\documentclass[10pt,onecolumn,a4paper]{article}
\newcommand{\authormark}[1]{\textsuperscript{#1}}
\usepackage{wasysym}
\usepackage{graphicx}
\usepackage{natbib}
\usepackage{epstopdf}
\begin{document}

\begin{center}
	\textbf{\LARGE The LArase Satellites Spin mOdel Solutions (LASSOS):  a comprehensive model for the spin evolution of the LAGEOS and LARES satellites \\[1.5em]}
	\large M. Visco \authormark{1,2},
	D. M. Lucchesi\authormark{1,2,3}\\[1em]
	\normalsize
		\authormark{1}IAPS - INAF , Via del Fosso del Cavaliere, 100, 00133 Roma, Italy \\
		\authormark{2} INFN, Sez. Tor Vergata, Via della Ricerca Scientifica 1, 00133  Roma, Italy \\ 
		\authormark{3} ISTI - CNR,  via G. Moruzzi 1, 56124 Pisa, Italy \\ 
		[3em]
\end{center}         

\begin{abstract}
The two LAGEOS and LARES are laser-ranged satellites tracked with the best accuracy ever achieved. Using their range measurements many geophysical parameters were calculated and some General Relativity effects were directly observed. To obtain precise and refined measurements of the effects due to the predictions of General Relativity on the orbit of these satellites,  it is mandatory to model with high precision and accuracy all other forces, reducing the free parameters introduced in the orbit determination. A main category of non-gravitational forces to be considered are those of thermal origin, whose fine modeling strongly depends on the knowledge of the evolution of the spin vector.  
We present a complete model, named LASSOS, to describe the evolution of the spin of the LAGEOS and LARES satellites. In particular, we solved Euler equations of motion considering not-averaged torques. This is the most general case, and the  predictions of the model well fit the available observations of the satellites spin. 
We also present the predictions of our model in the fast-spin limit, based on the application of averaged equations. The results are in good agreement with those already published, but with our approach we have been able to highlight small errors within these previous works. 
LASSOS was developed within the LARASE research program. LARASE aims to improve the dynamical model of the two LAGEOS and LARES satellites to provide very precise and accurate measurements of relativistic effects on their orbit, and also to bring benefits to geophysics and space geodesy. 
\end{abstract}

\section{Introduction}
The two LAGEOS (LAser GEOdynamics Satellite) and LARES (LAser RElativity Satellite) are passive Earth orbiting satellites. They are almost spherical in shape, with a diameter of about 60 cm and mass of about 407 kg and 405 kg, respectively for LAGEOS and LAGEOS II, and a diameter of 36.4 cm and a mass of about 387 kg in the case of LARES.   A large number of retroreflectors,  the so-called cube-corner retro-reflectors (CCRs), is distributed in the form of rings over the satellites' surface. These CCRs allow a precise tracking of the orbit of the satellites by means of a network of Earth bound stations coordinated by the International Laser Ranging Service (ILRS), see \cite{2002AdSpR..30..135P}. In particular, the two LAGEOS have embedded in their surface 426 CCRs, 422 made of fused silica and 4 of germanium, while the smaller LARES has 92 CCRs, all made of fused silica.

The older LAGEOS was launched by NASA in 1976, on May 14. LAGEOS has an orbit inclination of about $109.9^{\circ}$ over the Earth's equator, a semi-major axis of about 12,270 km and an eccentricity of about 0.004.
LAGEOS II was launched by NASA/ASI in 1992, on October 22. LAGEOS II has an orbit inclination of about $52.7^{\circ}$ over the Earth's equator, a semi-major axis of about 12,162 km and an eccentricity of about 0.014. The orbital periods of the two satellites are respectively of 225.4~min and  222.4~min. LARES was launched by ASI in 2012, on February 13. LARES has an inclination of $69.5^{\circ}$ over the Earth's equator, a semi-major axis of about 7,820~km, an eccentricity close to zero and an orbital period of 114.9~min.

The ground stations of the ILRS assure, by means of the Satellite Laser Ranging (SLR) technique, almost continuous measurements of the round trip time of narrow laser pulses and, consequently, of the satellites' distance, which is usually called range. High quality range data, called Normal Points (NPs), are produced averaging the tracking measurements over appropriate time spans, see \cite{Sinclair2012}.

The orbital parameters, obtained from the NPs with least squares methods, can be compared with those calculated using parametric dynamical models, allowing on one side an accurate evaluation of many geophysical parameters (see for instance \cite{1983Natur.303..757Y},
\cite{1984JGR....89.1077R}, \cite{1985JGR....90.9217C}, \cite{1990JGR....9522013S}, \cite{1998Lemoine}, \cite{1998P&SS...46.1633B}, \cite{2002Sci...297..831C} and \cite{2013GeoRL..40.2625C}) and, on the other side, a direct measurement of the relativistic gravitational effects acting on the satellite orbit, see for instance \cite{1996NCimA.109..575C}, \cite{2004Natur.431..958C}, \cite{2007AdSpR..39..324L}, \cite{2010PhRvL.105w1103L} and \cite{2014PhRvD..89h2002L}. 

The dynamical model used to calculate the orbit of these geodetic satellites must include also forces related with gravitational and non-gravitational perturbations (NGP). A main disturbing role is played by the surface forces strictly connected with the evolution of the satellites' spin vector (orientation and rate), we refer to \cite{1991JGR....96.2431B}, \cite{1994PhRvD..50.6068H}, \cite{1996JGR...10117861F}, \cite{1996GeoRL..23.3079V}, \cite{2007Andres} for details. These are the Earth-Yarkovsky and Yarkovsky-Schach thermal effects (see \cite{1987JGR....9211662R}, \cite{1988JGR....9313805R}, \cite{1989AnGeo...7..501A}, \cite{1990A&A...234..546F}, \cite{1991JGR....96..729S}, \cite{1996CeMDA..66..131S}, \cite{1996P&SS...44.1551F}, \cite{1997JGR...102..585R}, \cite{1997JGR...102.2711M}, \cite{1999AdSpR..23..721M}, \cite{2002P&SS...50.1067L}, \cite{2004P&SS...52..699L}, \cite{2006Andres}). 

The two LAGEOS satellites and LARES were injected in their orbit with an initial rotation that slows down in time under the magnetic torque, and with an initial orientation of their spin that were subject to an evolution from the precession due to the combined action of three main forces: magnetic, gravitational and surface forces. Several spin models were introduced in the past. Among these models, those that better describe the experimental data are the ones valid in the so called \textit{fast-spin} approximation.
In this approximation, the solutions of the models are valid as long as the satellite rotation period is much shorter than the orbital one. 
Nowadays, while LARES spins with a period of about 1000~s, one order of magnitude smaller then the orbital period, the two LAGEOS' spin periods are, respectively, about $3\cdot 10^5$ s for the older satellite and about $3\cdot 10^4$ s for the younger one. Both satellites spin less than one revolution during each orbit, therefore a more general model is needed beyond the \textit{fast-spin} approximation.

The aim of the present work is to introduce our spin model. This model has been developed within the activities of the LAser RAnged  Satellites Experiment (LARASE) \cite{2015CQGra..32o5012L,7573270}. The main new characteristic of our model is its capacity to simulate completely the spin trend of LAGEOS and LARES satellites in the general case, with no restriction to the mentioned \textit{fast-spin} approximation. In other words, the model is valid for any rotational period of the satellite. After a brief summary about the models developed in the past (see section \ref{old_mod}), in section \ref{our_sol} we focus on the equations we used and the different torques we considered.  Starting from these formulas we have been also able to calculate the  solution valid for averaged equations in the limit of the \textit{fast-spin} approximation. In section \ref{solut} we describe the numerical integration methods used to solve the equations in the general case and we compare the obtained results with the available experimental observations.
In section \ref{dis} we provide a discussion on the importance of a refined modeling for the spin evolution, and we also focus on the order of magnitude of other torques that are however negligible with respect to those taken into account in previous sections.
Finally, in section \ref{concl} our conclusions and recommendations are provided together with the work needed to further improve the free parameters that characterize our model for the spin of LAGEOS-like satellites.

\section{Previous models for the spin of the two LAGEOS satellites} \label{old_mod}
The problem of the comprehension of the rotational behavior of the older of the two LAGEOS satellites was first studied by \cite{1991JGR....96.2431B}.  In that paper the authors, in order to model the spin of LAGEOS, considered two torques acting over the satellite. The first torque is the one produced by the interaction of the magnetic moment  of the satellite with the Earth's Geomagnetic field. The magnetic moment is the one produced by the eddy currents (or Foucault currents) induced in the conductive body of the satellite from its rotation and motion in the same field. The second torque arises from the action of the Earth gravitational field, because of the non spherical mass distribution of the satellite. 

\cite{1991JGR....96.2431B} solved the problem in the \textit{fast-spin} approximation by averaging the torques on the satellite orbital period and on the Earth rotational period.  
In that work for the first time the measured decay of the spin of LAGEOS  was explained with a good accuracy and, at the same time, the complexity and variety of the results in different regions of the space of parameters were shown.
An additional significant result of \cite{1991JGR....96.2431B} model was that, with respect to the inversion of the initial sense of rotation of the satellite, the evolution of both the direction and rate of the spin are not invariant.

The analysis of the spin model of LAGEOS --- as well as the results of this first successfully work --- were successively extended by \cite{1996JGR...10117861F} and \cite{1996GeoRL..23.3079V}.  In \cite{1996JGR...10117861F}, some small errors present in the formulas published by \cite{1991JGR....96.2431B} were corrected and, most importantly, these authors extended the model in order to take in account of a possible misalignment between the spin and symmetry axes. They also showed that the \cite{1991JGR....96.2431B} spin model leads to a successful fit of the along--track residuals of the  two LAGEOS related with the orbital perturbation produced by the main thermal effects, once the force mechanisms related with the direction of the satellite spin vector were properly modeled. 

Finally, \cite{1996JGR...10117861F} considered other possible torques connected to atmospheric skin forces and radiation pressure: one first torque connected to the difference of reflectivity between the two LAGEOS's hemisphere \cite{1991JGR....96..729S} and the second arising from the non coincidence between the geometric center and the center of mass of the satellite. However, they wrongly concluded that likely these torques would not play an important role in the evolution of the spin of the two LAGEOS satellites, 
because negligible and that, at first order, they should be ignored. We refer to \cite{1996GeoRL..23.3079V} for details. 
This conclusion was probably dictated by the fact that these authors have not tuned their spin model with the available (at that time) observations of the spin of the two LAGEOS satellites, but, conversely, with the residuals in their semi-major axis.

These two (much smaller) torques were inserted later in the LOSSAM  (LageOS Spin Axis Model) model (see \cite{2004JGRB..10906403A, 2007Andres}), that till now was the most complete and reliable model to describe the spin behavior of LAGEOS satellites in the \textit{fast-spin} approximation (\cite{2013AdSpR..52.1332K}) and recently it was also applied with success to LARES, see \cite{6575147}.
Indeed, the LOSSAM model has represented a huge improvement in  modelling the evolution of the spin of the two LAGEOS satellites with respect to previous models, even if the bulk of the characteristics of the torques considered are exactly the one previously developed by \cite{1991JGR....96.2431B} --- for the magnetic and gravitational torques --- and by \cite{1996JGR...10117861F} for what concerns the additional (minor) torques related with the radiation pressure asymmetry and the possible offset between the satellites geometric center and their center of mass.

The key aspects that have led to the development of a so successful model can be summarized in the following main points:
\begin{enumerate}
	\item{
		the collection of all available observations (up to 2007) of the spin of the two LAGEOS satellites, both in their rate and orientation, see \cite{1980SPIE..227..148S}, \cite{1997PhDT........14A}, \cite{GRL:GRL14288} and \cite{2004ITGRS..42..202O};}
	\item{a careful analysis of all the parameters that enter in the models, i.e., in the mathematical expressions of the various torques;}
	\item{the tuning of a suitable subset of these parameters by means of a least-squares fit of the LOSSAM model to all the available observations.}
\end{enumerate}

It is important to underline that, in order to analyze the spin evolution under the most general conditions, the torques expressions used in the equations should not be averaged. In this case it is handy to work in the rotating frame of the satellite and use the Euler's equations. \cite{1994PhRvD..50.6068H} were the first to formulate the problem in this way. Their aim was to qualitatively analyze the time evolution of the satellites rotation, and they did not care if their solution did not fit to the experimental data. Anyway, their model was too much simplified to replicate the measured data because of several assumptions. In fact: i) the expression adopted for the polarizability of the satellite is too far from reality, ii) the expression of the magnetic torque is the same of \cite{1991JGR....96.2431B} and \cite{1996JGR...10117861F}, that is valid only in the \textit{fast-spin} approximation, iii) the Earth magnetic field is simplified as a pure dipole field along the Earth rotation axis, finally iv) the equations of motion are written in a non inertial reference frame, neglecting the precession of the orbital plane with respect to an inertial reference frame related to the Earth.

However, despite of these excessive oversimplifications, \cite{1994PhRvD..50.6068H} have correctly introduced the general approach to be followed and based on the resolution of the Euler dynamical equations, and they also introduced the correct general expression for the rate of precession in the case of the gravitational torque, correcting the expression given by \cite{1991JGR....96.2431B} (that was anyway used later by \cite{1996JGR...10117861F}).

The  \cite{1994PhRvD..50.6068H} approach based on the numerical solution of the full set of the Euler equations was resumed a few years later by \cite{2002PhDT.......176W}, removing the simplifications of the previous model but, at the end, still with an unsuccessful fit to the available observational data. 
Later on, this more general model was also considered by \cite{2002EGSGA..27.2994A} and \cite{2003EAEJA.....5951A}, that developed the so called ''real-time'' LOSSAM formulation.  We could not find in literature any example of practical application of these models. In fact, \cite{2007Andres} wrote in his PhD thesis work that 
 ``real-time'' LOSSAM had not given a \textit{better result} than the model developed by \cite{2002PhDT.......176W}, giving implicitly a negative judgment on the results obtained by both models.   

\section{The LASSOS model for the evolution of the spin}
\label{our_sol}
In the following we present our new model for the spin of LAGEOS and LARES satellites, that aims to be general and complete. All the main known torques were considered and their mathematical expressions were written in a general (not averaged) way in the rotating frame of the satellite adopting the Euler's rotational equations.

\subsection{Reference frames}
For our analysis, in the following, we will adopt two different reference frames: 
\begin{itemize}
	\item the Earth Mean Equator and Equinox of Date Frame (J2000)
	\item the satellite Body Frame (BF).
\end{itemize}

The J2000 is identified by three axes ($ {\bf \hat{x}^E, \hat{y}^E, \hat{z}^E}$). The origin of this Cartesian reference frames coincides with the Earth's center, the $ {\bf (\hat{x}^E,\hat{y}^E)}$ plane coincides with the Earth's mean equatorial plane. The $ {\bf \hat{x}^E}$  axis is directed to the vernal equinox $\vernal$, intersection between the mean equator and the mean equinox of date. 
The J2000 represents a quasi-inertial frame.

The BF (also known as Body-Fixed reference frame) is centered in the body's center of mass with the axes ($ {\bf \hat{x}^b, \hat{y}^b ,\hat{z}^b} $) aligned along the principal axes of inertia. 
The position of the BF can be conveniently expressed with respect to the J2000 reference frame as function of the three Euler's angles $\theta$, $\phi$ and $\psi$. 
The nutation angle $\theta$ is the angle between $ {\bf \hat{z}^E}$ and $ {\bf \hat{z}^b}$, the spin angle $\psi$ is the angle between the nodal line and $ {\bf \hat{x}^b}$ axis, the precession angle $\phi$ is the angle between   $ {\bf \hat{x}^E}$ axis and the nodal line, where the nodal line is the intersection between the two planes $ {\bf (\hat{x}^E,\hat{y}^E)}$ and $ {\bf (\hat{x}^b,\hat{y}^b)}$. 
The transformations of a vector  ${\bf V}$ from J2000 to BF reference frame  ${\bf V^b =\mathcal{R} \; V^E}$  is given by the rotation matrix ${\bf \mathcal{R}}$, that in terms of Euler's angles is (\cite{Goldstein}):

	\begin{eqnarray}
	{\bf \mathcal{R}}=\left(
	\begin{array}{c c c}
	\cos\phi~\cos\psi - \cos\theta~\sin\phi~\sin\psi&\quad \sin\phi~\cos\psi + \cos\theta~\cos\phi~\sin\psi&\quad \sin\theta~\sin\psi\\
	- \cos\phi~\sin\psi - \cos\theta~\sin\phi~\cos\psi&\quad \cos\theta~\cos\phi~\cos\psi - \sin\phi~\sin\psi&\quad \sin\theta~\cos\psi\\
	\sin\theta~\sin\phi& - \cos\phi~\sin\theta& \cos\theta
	\end{array}
	\right).		
	\label{Eu_matrix}
	\end{eqnarray}

The satellite motion along its orbit, supposed to be quasi-circular with radius $a$, will be identified by the Keplerian orbital parameters measured in the J2000 frame: $\Omega$, the right ascension of the ascending node, $I$, the orbit inclination over the Earth's equator, $\omega$, the argument of pericenter, and $M_0$, the mean anomaly.

\subsection{Equations of motion}
The spin evolution can be conveniently described by the Euler's equations

\begin{eqnarray}
I_x~\dot{\omega}_{sx}^b-\omega_{sy}^b \omega_{sz}^b (I_y-I_z)=M_x \nonumber\\
I_y~\dot{\omega}_{sy}^b-\omega_{sx}^b \omega_{sz}^b (I_z-I_x)=M_y\\
I_z~\dot{\omega}_{sz}^b-\omega_{sx}^b \omega_{sy}^b (I_x-I_y)=M_z \nonumber
\label{Eu_eq}
\end{eqnarray} 

where $I_x,I_y,I_z$  are the moments of inertia with respect to the principal axes of inertia of the satellite body, while $M_x, M_y $ and $M_z$ are the components of total torque $ {\bf M}$ along the same axes. 
The components of the angular velocity measured with respect to the body axes, $\omega_{sx}^b, \omega_{sy}^b$ and $\omega_{sz}^b$ can be substituted by their expressions in term of the Euler angles in the J2000 reference frame \cite[see][]{Goldstein}:
\begin{eqnarray}
\omega_{sx}^b&=&\dot{\phi}~\sin\theta \sin\psi+ \dot{\theta} ~\cos\psi \nonumber \\
\omega_{sy}^b&=&\dot{\phi}~\sin\theta \cos\psi- \dot{\theta} ~\sin\psi \nonumber \\
\omega_{sz}^b&=&\dot{\phi} ~\cos\theta + \dot{\psi}.
\label{omgex}
\end{eqnarray}

We finally obtain the Euler's equations in term of Euler's angles measured in the J2000 reference frame:

\begin{eqnarray}
	\ddot{\theta}&=&\frac{M_x}{ I_x}\cos\psi - \frac{M_y}{ I_y}\sin\psi - \frac{I_z}{I_y} \dot{\phi} \dot{\psi} \sin \theta  +\frac{I_y-I_z}{I_x} \dot{\phi}^2 \frac{\sin(2 \theta)}{2}
	\nonumber \\
	&&\label{Eu_eq1} +\frac{I_x-I_y}{I_x} \frac{\Lambda}{ I_y} \left[\dot{\theta} \left(\dot{\psi}+\dot{\phi} \cos \theta  \right)  \frac{\sin(2 \psi)}{2} +\dot{\phi}^2 \frac{\sin(2 \theta)}{2} \sin^2 \psi - \dot{\phi} \dot{\psi} \sin \theta  \left( \frac{ I_y -  I_z}{\Lambda} - \sin^2 \psi \right) \right] \\
	\nonumber \\
	\ddot{\phi}&=& \frac{M_y}{I_y} \frac{\cos \psi}{\sin \theta}+ \frac{M_x}{I_x}\frac{ \sin \psi}{\sin \theta} +\frac{I_z}{I_y} \frac{\dot{\psi} \dot{\theta}}{ \sin \theta} - \frac{\Lambda}{I_x} \frac{\cos \theta}{\sin \theta}\dot{\phi} \dot{\theta}  + \nonumber \\ 
	&&\label{Eu_eq2} \frac{I_x-I_y}{I_y} \frac{\Lambda}{I_x}\left[ \frac{1}{ \sin \theta }\left( \sin^2 \psi - \frac{I_x}{\Lambda}\right)\dot{\psi} \dot{\theta}
	-  \frac{\sin(2 \psi)}{2} \left( \cos \theta \dot{\phi} + \dot{\psi} \right)  \dot{\phi} 
	- \frac{\cos \theta}{\sin \theta } \dot{\phi} \dot{\theta}\cos^2 \psi \right]  \\ 
	\nonumber \\
	\label{Eu_eq3} \ddot{\psi}&=&\frac{M_z}{I_z} - \frac{\cos(\theta)}{\sin(\theta) } \left(\frac{M_y}{I_y}\cos(\psi) +\frac{ M_x}{I_x}\sin(\psi)\right) + \dot{\phi} \dot{\theta}\frac{1}{\sin \theta} \left(\frac{ I_y -  I_z}{ I_x }\cos^2 \theta + 1\right) - \frac{ I_z}{ I_y } \dot{\psi} \dot{\theta} \frac{\cos \theta}{\sin \theta }+ \nonumber\\ 
	&& \left( I_x -  I_y \right) \left[\frac{1}{ I_z} \dot{\phi} \dot{\theta}  \frac{1}{\sin \theta}\left(\sin^2 \theta \cos(2 \psi)+ \frac{ \Lambda}{ I_x  I_y}\cos^2 \psi \cos^2 \theta \right)- \dot{\theta}^2 \frac{1}{2I_z}\sin(2\psi)-\dot{\phi}^2\frac{1}{2I_z}\sin(2 \psi) \right. \nonumber \\ &&\left.\left(\frac{\Lambda I_z}{ I_x  I_y}\cos^2 \theta -\sin^2 \theta \right)  - \dot{\psi} \dot{\theta} \frac{1} { I_y}\frac{\cos \theta}{\sin(\theta)}  \left(\frac{\Lambda}{I_x}  \sin^2 \psi - 1 \right)+ \frac{ \Lambda}{2 I_x  I_y} \dot{\phi} \dot{\psi} \cos \theta \sin(2 \psi) \right],
	\end{eqnarray}
	where $\Lambda= I_x+I_y -  I_z$.\\

These equations were written without hypothesizing any peculiar symmetry for the satellite, that is we assumed $I_x \neq I_y \neq I_z$ differently from \cite{1994PhRvD..50.6068H}, who instead have adopted an axial symmetry for the satellite. 
Coherently,  our expressions (\ref{Eu_eq1}), (\ref{Eu_eq2}) and (\ref{Eu_eq3}) converge, in the limit of $I_x=I_y$ (cylindrical symmetry), to expressions (9)-(14) of \cite{1994PhRvD..50.6068H}. 
We underline that even if in this limit the formal expressions are equal, our Euler angles are expressed in the J2000 reference, while \cite{1994PhRvD..50.6068H} adopted as main reference frame the orbital one.

In our analysis, for the torques to be included in equations (\ref{Eu_eq1})-(\ref{Eu_eq3}) we considered four contributions: 

\begin{itemize}
	\item{$ {\bf M}_{mag}$, i.e. the torque from the Earth magnetic field (section \ref{Mag_field})}
	\item{$ {\bf M}_{grav}$, i.e. the torque from the Earth gravitational field  (section \ref{Grav_field})}
	\item{$ {\bf M}_{off}$, i.e. the torque  arising from radiation pressure acting on the geometric center of the satellite,  if it does not coincide with its center of mass (section \ref{cm_diff})}
	\item{$ {\bf M}_{asy}$, i.e. the torque from radiation pressure acting on the two opposite hemispheres of the two LAGEOS satellites, which have a different reflectivity (section \ref{ax_rad})}
\end{itemize}

therefore 
$ {\bf M}= {\bf M}_{mag}+ {\bf M}_{grav}+ {\bf M}_{off}+ {\bf M}_{asy}$.

\subsection{Torques in time domain} \label{Gen_sol}
In the following sub-sections we will calculate the expressions of the main torques, to be inserted in previous equations (\ref{Eu_eq1}), (\ref{Eu_eq2}) and (\ref{Eu_eq3}), as function of time. 

\subsubsection{Torque from the Earth magnetic field}\label{Mag_field}
Both LAGEOS satellites (\cite{1985JGR....90.9217C,2016AdSpR..57.1928V}) and LARES (\cite{2013AcAau..91..313P}) are made of conductive material, and even if they are supposed to be uncharged, they assume a magnetic moment while they are spinning inside the Earth magnetic field, the value and the direction of which change along their orbit. The induced magnetic moments $\textbf{m}$, because of the Earth magnetic field \textbf{B}, produce a torque on the satellites ($\textbf{M}_{mag}=\textbf{m}\times\textbf{B}$).

In the previously quoted papers, to describe such a phenomena, the LAGEOS satellite was modeled (\cite{1991JGR....96.2431B, 1994PhRvD..50.6068H, 1996JGR...10117861F}) as a conductive sphere rotating in a static magnetic field. The value of this constant magnetic field was calculated averaging the magnetic field of the Earth over the entire orbit of the satellite. The solution to this problem is well known in the literature, see for instance \cite{landau_1960}.   

This solution, which is completely valid in a quasi-stationary field, can be suitably used as long as the the rotation period of the satellite is much shorter than its orbital period as well as of the Earth's rotation period, but it could produce wrong results when is used in \textit{slow-spin} conditions, as in \cite{1994PhRvD..50.6068H}. 
These authors have probably adopted this simplification due to the lack of an handy model for a sphere rotating in an alternating magnetic field. The only solution apparently available till now in the literature is the one by \cite{hayes_1964} that unfortunately it is not easy to deal with.

In order to obtain a more general expression of the magnetic torque we faced the problem to find an easily integrable expression for the torque acting on a conducting sphere rotating in an alternating magnetic field. 
We found an expression for the torque that was applied to the satellites (modeled like perfect conducting spheres) while rotating and moving along their orbit:

 {\small 
	\begin{eqnarray}
	\label{eq_mag}
	\begin{array}{lll}
	 {\bf M}_{mag}^{E}=& V \sum^{8}_{i=0}    \frac{\left| {\bf B_i }\right|^2}{2 \left| {\bf \omega_s}\right|} \left\{-A''_i\left[1+\cos(2 \omega_i t+2 \varphi_{i}) \right]+D'_i \sin(2 \omega_i t+2 \varphi_{i}) \right\} {\bf \omega_{s}^{E}}+ \\
	\\
	&V  \sum^{8}_{i=0}\frac{  {\bf B_i} \cdot  {\bf \omega_s}}{2 \left| {\bf \omega_s } \right|^2} \left\{ \left[\alpha'(\omega_i) -A'_i\right] \left[1+\cos(2 \omega_i t+2 \varphi_{i}) \right]-\left[D''_i+\alpha''(\omega_i) \right] \sin(2 \omega_i t+2 \varphi_{i}) \right\} \left( {\bf \omega_s^{E}} \times  {\bf B_i}\right) + \\ 
	\\
	& V  \sum^{8}_{i=0}  \frac{  {\bf B_i} \cdot  {\bf \omega_s}}{2 \left| {\bf \omega_s }\right|}  \left\{A''_i \left[1+\cos(2 \omega_i t+2 \varphi_{i}) \right]-D'_i \sin(2 \omega_i t+2 \varphi_{i}) \right\}  {\bf B_i}, 
	\label{tor_nmed}
	\end{array} 
	\end{eqnarray}
	\begin{eqnarray}
	\begin{array}{lll}
	\mbox{where}  & & \\ 
	A'_i=\frac{\alpha'(\omega_s^E -\omega_i) +\alpha'(\omega_s^E + \omega_i)}{2} &\ \ \ \ \ & D'_i=\frac{\alpha'(\omega_s^E -\omega_i)-\alpha'(\omega_s^E + \omega_i)}{2} \\
	\\
	A''_i=\frac{\alpha''(\omega_s^E -\omega_i) +\alpha''(\omega_s^E + \omega_i)}{2}&\ \ \ \ &D''_i=\frac{\alpha''(\omega_s^E -\omega_i)-\alpha''(\omega_s^E + \omega_i)}{2}. \\
	\label{AD}
	\end{array}
	\end{eqnarray}
}

In the above expressions, 
$\alpha \left(\omega \right)=\alpha'+j\alpha''$ is the complex fourier transform of the magnetic polarizabilty per unity of volume of the satellite, $V$ is the satellite volume,  $\omega_i$ are the angular velocities of the harmonic components of the magnetic field, while $\bf{\omega}_s^E$ is the satellite spin angular velocity in J2000 frame, whose components can be expressed in term of the Euler angles \cite[see][]{Goldstein}:

\begin{eqnarray}
\omega_{s\ x}^E&=& \dot{\theta} ~\cos\phi+\dot{\psi}~\sin\theta \sin\phi \nonumber \\
\omega_{s \  y}^E&=&\dot{\theta} ~\sin\phi-\dot{\psi}~\sin\theta \cos\phi \nonumber \\
\omega_{s \  z}^E&=&\dot{\psi} ~\cos\theta + \dot{\phi}. 
\label{omgex2}
\end{eqnarray} In appendix A we calculated  the magnetic field experienced by the satellite as function of time along its orbit using the dipole approximation for the Earth's magnetic field (see appendix A for explicit expressions of $ {\bf B_i}$):

\begin{eqnarray}
 {\bf B}&=& \sum^{8}_{i=0}  {\bf B_i} \cos(\omega_i t+\varphi_{i}), 
\end{eqnarray}
where
\begin{eqnarray}
\label{freq_mag}
\omega_0&=&0 \nonumber\\
\omega_1&=&\omega_2=\omega_{\oplus}-2 n  \hspace{30pt} \nonumber\\
\omega_3&=&\omega_4=\omega_\oplus+2 n \hspace{30pt}  \\
\omega_5&=&\omega_6=2 n \hspace{30pt}  \nonumber \\
\omega_7&=&\omega_8=\omega_\oplus, \nonumber
\end{eqnarray}
and
\begin{eqnarray}
\varphi_{i}&=&\left\{
\begin{array}{cc}
-\frac{\pi}{2}& \mbox{for i=2,4,6,8} \nonumber \\
0& \mbox{for i=0,1,3,5,7}.
\end{array} 
\right.
\end{eqnarray}

In the previous expressions $\omega_\oplus$ represents the Earth's rotational angular velocity, while $n$ represents the satellite mean motion in the hypotheses of a circular orbit having radius $a$:

\begin{equation}
 {\bf n}=\sqrt{\frac{G M_{\oplus}}{a^3}}(\sin I \sin \Omega, -\sin I\cos \Omega, \cos I),
\end{equation} 

where $M_{\oplus}$ and $G$ are, respectively, the Earth's mass and the gravitational constant.

The torque that originated from Earth magnetic field
(\ref{eq_mag}) shows components at the magnetic field harmonics (\ref{freq_mag}) and their multiples. The torque is expressed as sum of several contributions: i) along the satellite angular speed direction (${\bf \omega_s^E}$), ii) along the magnetic field directions $ {\bf B_i}$  and iii) along directions orthogonal to the previous ones $\left( {\bf \omega_s^E} \times  {\bf B_i}\right) $. 

The expression (\ref{tor_nmed}) is calculated in J2000 frame, the matrix ${\bf \mathcal{R}}$ transforms it into the equivalent expression in body frame, to be used in the equations of motion (\ref{Eu_eq}): $ {\bf M_{mag}^b}={\bf \mathcal{R}}\   {\bf M}_{mag}^{E}$.  We have not determined an analytic expression for $ {\bf M_{mag}^b}$, because it is calculated numerically while solving the Euler's equations. 
The complex fourier transform  of the magnetic polarizability of the satellite $\alpha \left(\omega \right)$ can be approximated with the expression valid for a perfect sphere: we adopted the one included in \cite{CMR:CMR20090} calculated for a sphere having radius $R$, electrical conductivity $\sigma$, in a magnetic uniform field changing with an angular frequency $\omega$:

	\begin{eqnarray}
	\label{alfa}
	\alpha(\omega)&=&\alpha'+j \alpha''= \nonumber\\
	&=&\frac{3}{8 \pi}\left\{\frac{2 \mu_r \left[1-k \cdot \cot(k)\right]+\left[1-k^2-k \cdot \cot(k)\right]}{\mu_r \left[1-k \cdot \cot(k)\right]-\left[1-k^2 -k \cdot \cot(k) \right]}\right\},
	\end{eqnarray}
	where
	\begin{eqnarray}
	k(\omega)=\frac{R}{\delta(\omega)} \left(1+j\right), \\  
	\delta(\omega)=\frac{c}{\sqrt{2 \pi \omega \sigma \mu_r }}.
 \end{eqnarray}
This expression contains the dependence from  the relative magnetic permeability $\mu_r$, that was  ignored  by all the other authors (\cite{1991JGR....96.2431B, 1994PhRvD..50.6068H, 1996JGR...10117861F, 2007Andres}). Even if $\mu_r$ can be very close to 1 (for instance for aluminum $\mu_r-1\cong2.2\cdot 10^{-5}$), its contribution could be non negligible at very low frequencies, where it represents the residual magnetic polarizability of a non-spinning satellite.  

Past experience with the two LAGEOS  (\cite{1991JGR....96.2431B, 1996JGR...10117861F, 2007Andres, 2012AdSpR..50.1473K}) has shown that the calculated values of the time evolution of the spin of LAGEOS satellites fit well the experimental data only if a low frequency approximation is chosen for  (\ref{alfa}):

{\small 	\begin{eqnarray}
	\label{red_alfa}
	\alpha(\omega)&=&\alpha'+j \alpha''= 
\\&\simeq & \left[\frac{3}{4~\pi} \frac{\mu_r-1}{\mu_r+2} -\frac{9}{350 \pi} \frac{\mu_r \left(\mu_r+9\right)}{\left(\mu_r+2\right)^3}\left(\frac{R}{\delta}\right)^4 \right]\beta'  +j \frac{9}{20 \pi}\frac{\mu_r}{\left(\mu_r+2\right)^2}\left(\frac{R}{\delta}\right)^2 \beta'',\nonumber \label{alfas}
	\end{eqnarray}}

$\beta'$ and $\beta''$ are dimensionless constants to be determined experimentally in order to take in account the differences of the satellite's shape  with respect to an ideal sphere. 

There is no trivial explanation to justify such a simplified empirical model. It was hypothesized that the composite structure of LAGEOS, an inner cylindrical  core made of brass inserted in two hollow aluminum hemispheres, would invalidate the assumption of a uniform sphere as a model for LAGEOS. We will devote a forthcoming paper to discuss in detail this case that, for now, we adopt as a matter of fact. In the case of LARES, made of a unique block of metal, we believe that it is more correct to apply the expression (\ref{alfa}).

The first term of the real part of (\ref{red_alfa}) is introduced for the first time in this work, and it is an important term, because it represents the residual polarizability at zero frequency. This term dominates the real part of the polarizability at frequencies lower than $\omega_{c}$:

\begin{eqnarray}
\omega_{c}=\frac{5 c^2 }{12 \pi R^2 \sigma } \frac{\left(\mu_r+2\right) }{\mu_r}\sqrt{\frac{42\left(\mu_r-1\right)}{ \mu_r \left(\mu_r+9 \right)}}.
\end{eqnarray}

This bound frequency corresponds for the two LAGEOS  to a period of about  $T_c=2 \pi/\omega_{c}\sim400~s$, that was already reached by both satellites.

We underline that there is a big difference between the formula we have used for the torque of magnetic origin (\ref{eq_mag}), and that valid for static magnetic fields used till now by the others authors (see for instance (2), (5) and (6) of \cite{1996JGR...10117861F}).  
In formula (\ref{eq_mag}), the torque along the satellite angular velocity ${\bf \omega_s}$ could change sign when 
$ \omega_s= \omega_i$, depending on  $A_i''$ that switch from negative to positive. In this case the angular velocity $\omega_s$ could be maintained equal to ${\bf \omega_i}$ by the magnetic force. 
This resonant condition, however, appears only if the ratio $\frac{R} {\delta (\omega_i)} $ is greater than a threshold value that is not reached anyway by LAGEOS or LARES. A more detailed discussion around these very important issues will be included in a forthcoming work.

We point out that two possible exponents can be used in the Fourier transform  (i.e.  $j\omega$ or $-j \omega$) and, consequently, we can have two different expressions with different complex conjugate for $\alpha$. One must be careful to maintain the same convention in the inversion formula in order to anti-transform in the time domain.

\subsubsection{Torque from the Earth Gravitational field}\label{Grav_field}
In this section we  calculate the components of the torque of gravitational origin $M_{grav}$, also known as gravity gradient torque, to be included in equations (\ref{Eu_eq1}), (\ref{Eu_eq2}) and (\ref{Eu_eq3}). In fact, because the satellites do not have a perfect spherical symmetry in their mass distribution, the Earth gravitational field produces a torque on them. We restrict our analysis to the monopole component of this field, whose general expression is given by (\cite{Bele}):

{\small 	\begin{equation}
	\label{gravity}
	 {\bf M}_{grav}^b=3 n^2 \left\{ ~ {\bf \hat{s}^b}\times \left[ I_x \left( {\bf \hat{s}^b} \cdot  {\bf \hat{x}^b} \right) {\bf\hat{x}^b}  +I_y \left(  {\bf \hat{s}^b} \cdot  {\bf \hat{y}^b} \right) {\bf \hat{y}^b} +I_z \left( {\bf \hat{s}^b}  \cdot  {\bf \hat{z}^b} \right) {\bf\hat{z}^b} \right] \right\} =3 n^2
	\left[
	\begin{array}{c}
	\left(I_y-I_z\right) s_{y}^b~s_{z}^b\\
	\left(I_z-I_x\right) s_{x}^b~s_{z}^b\\
	\left(I_x-I_y \right)  s_{x}^b~s_{y}^b\\
	\end{array}	
	\right],
	\end{equation}}

where $ {\bf \hat{s}^b}$ is the unit vector Earth-to-satellite, having components $s_x^b, s_y^b, s_z^b$ in the BF reference frame. 

To be conveniently included in the equations of motion (\ref{Eu_eq1}-\ref{Eu_eq3}), (\ref{gravity}) must be expressed as a function of the Euler angles with respect to the J2000 reference frame.

The expression of the three components of the Earth-to-satellite unit vector can be  easily calculated in the J2000 reference frame (see for instance \cite{montenbruk}) and  can be further converted into the BF reference frame using (\ref{Eu_matrix}) for the rotation matrix $ {\bf \mathcal{R}}$, finally obtaining:

{\small 	\begin{eqnarray}
	\label{sxsysz}
	s_x^b&=&\cos(\omega+M_0+n \cdot t) \left[\cos( \phi- \Omega) \cos( \psi) -  \cos(\theta) \sin( \phi- \Omega) \sin( \psi) \right] + \nonumber \\
	&&\sin(\omega+M_0+n \cdot t) \left\{\sin( \phi- \Omega)   \cos(I) \cos( \psi)+\left[ \sin(\theta) \sin(I)+
	\cos(\theta) \cos( \phi- \Omega) \cos(I)  \right]  \sin(\psi)  \right\} \nonumber \\
	\nonumber \\
	s_y^b&=& - \cos(\omega+M_0+n \cdot t)\left[  \cos(\theta) \sin( \phi- \Omega) \cos( \psi) +\cos( \phi- \Omega) \sin( \psi) \right] + \\
	&& \sin(\omega+M_0+n \cdot t) \left\{ \left[  \sin(\theta)  \sin(I)+ \cos(\theta) \cos( \phi- \Omega)\cos(I) \right]  \cos( \psi)-\sin( \phi- \Omega) \cos(I) \sin( \psi)   \right\} \nonumber \\
	\nonumber \\
	s_z^b&=& \cos(\omega+M_0+n \cdot t) \sin(\theta) \sin( \phi- \Omega)+ \nonumber \\
	&&\sin(\omega+M_0+n \cdot t) \left[\cos(\theta) \sin(I) -
	\sin(\theta) \cos( \phi- \Omega) \cos(I) \right]. \nonumber
	\end{eqnarray}}

Obviously, from equation (\ref{gravity}) follows that the gravitational torque is zero if the satellite has a perfect spherical symmetry (i.e. $I_x=I_y=I_z$). Moreover, the unit vectors of Eqs. (\ref{sxsysz}) are periodic with the satellite mean motion frequency $n$, therefore their product in the expression (\ref{gravity}) for the gravitational torque produces a constant component plus one periodic component with a frequency $2n$.

\subsubsection{Offset torque} \label{cm_diff}
The radiation pressure on the satellite surface acts, at the end, on the geometric center of the satellite. If the geometric center does not coincide with the center of mass of the satellite, the radiation pressure produces a torque (\cite{1996GeoRL..23.3079V}). If the vector ${\bf h^b}$ identify the geometric center position respect to the center of mass in the body reference frame, the torque expression is:

\begin{eqnarray}
 {\bf M_{off}^b}&=&\nu \pi R^2~\frac{\Phi_{\odot}}{c}~C_R~\left( {\bf h^b \times} {\bf \hat{s}_\odot^b}\right).\label{eq_off}
\end{eqnarray}

In this equation $R$ is the radius of the satellite, $C_R$ represents its radiation coefficient (related with the reflectivity of its surface), $\Phi_{\odot}$ is the solar flux at the Earth-Sun distance, $ {\bf \hat{s}_\odot^b}$ is the satellite-to-Sun unity vector, finally $\nu$ (with $0\leq \nu\leq 1$) represents the shadow function. This function measures the percentage of solar flux that reaches the satellite during the eclipses by Earth (with $\nu=1$ when there is not an eclipses, and $\nu=0$ when the eclipse is complete; see for instance \cite{montenbruk}).  
The expression (\ref{eq_off}) is more general of the one by \cite{1996GeoRL..23.3079V}, that hypothesized the coincidence between the rotation and symmetry axes.

To include the torque coming from the offset (\ref{eq_off}) into the equations of motion  (\ref{Eu_eq1}-\ref{Eu_eq3}), the unit vector $ {\bf \hat{s}_\odot^b}$
must be expressed in the BF reference frame in term of the Euler angles with respect to the J2000 reference frame. 

If the motion of the Sun in the J2000 reference frame is considered circular with constant angular velocity $\omega_\odot$ along the ecliptic, by denoting with  $\epsilon$ the angle between the plane of the ecliptic and the Earth's equator, and with $\lambda_{\odot}$ the Sun's ecliptic longitude at t=0,  the three components $s_{\odot x}^b, s_{\odot y}^b$ and  $s_{_\odot z}^b$ of the  Earth-to-Sun unit vector in the BF reference frame are:

{\small 	\begin{eqnarray}
	s_{\odot x}^b&=& \sin(\lambda_{\odot} + \omega_\odot \cdot t) \left[ \cos(\epsilon) (\sin( \phi) \cos( \psi)   +  \cos( \theta)\cos( \phi)    \sin( \psi)) +  \sin(\epsilon)    \sin( \theta)\sin( \psi) \right] + \nonumber \\
	&&\cos(\lambda_{\odot} + \omega_\odot \cdot t) \left[ \cos( \phi)  \cos( \psi) -  \cos( \theta)  \sin( \phi)  \sin( \psi) \right] \nonumber\\
	s_{\odot y}^b&=&   -  \sin(\lambda_{\odot} + \omega_\odot \cdot t) \left[ \cos(\epsilon) ( \sin( \phi)  \sin( \psi) -  \cos( \phi)  \cos( \psi)  \cos( \theta)) -  \cos( \psi)  \sin(\epsilon)  \sin( \theta) \right] - \nonumber\\
	&& \cos(\lambda_{\odot} + \omega_\odot \cdot t) \left[ \cos( \phi)  \sin( \psi) +  \cos( \psi)  \cos( \theta)  \sin( \phi) \right] \\
	s_{\odot z}^b&=& \sin(\lambda_{\odot} + \omega_\odot \cdot t) \left[ \sin(\epsilon)  \cos( \theta) -  \cos(\epsilon)  \cos( \phi)  \sin( \theta) \right] +   \cos(\lambda_{\odot} + \omega_\odot \cdot t)  \sin( \theta)\sin( \phi). \nonumber
	\end{eqnarray}
 }

This torque, slowly changing with a period of one solar year, is modulated by the shadow function $\nu$.

\subsubsection{Anisotropic  reflection torque} \label{ax_rad}
A difference in the reflectivity of the two hemisphere of LAGEOS was hypothesized by \cite{1991JGR....96..729S} in order to explain the observed along-track residuals in acceleration of the satellite. This empirical result was reasonably explained by \cite{2003GeoRL..30.1957L,2004CeMDA..88..269L} as due by a possible asymmetry in the satellite's reflection of the visible solar radiation introduced by the four germanium CCRs. If the relative difference between the reflectivity of North and South hemispheres is $\Delta \rho=(C_R^N-C_R^S)/\bar{C}_R$, with  
$\bar{C}_R=(C_R^N+C_R^S)/2$, the following torque is produced (\cite{1996GeoRL..23.3079V}):

{\small 	\begin{eqnarray}
	\label{eq_asym}
	 {\bf M_{ar}^b}&=&\nu~\frac{2}{3}R^3~\frac{\Phi}{c}~\Delta \rho~C_R  ~( {\bf  \hat{z}^b \times \hat{s}_\odot^b })\left| {\bf \hat{z}^b \times \hat{s}_\odot^b }\right|=\nu~\frac{2}{3}R^3~\frac{\Phi}{c}\Delta \rho~C_R~\sqrt{\left(s_{\odot x}^b\right)^2+\left(s_{\odot y}^b\right)^2} \left[
	\begin{array}{c}
	-s_{\odot y}^b\\
	s_{\odot x}^b\\
	0
	\end{array}
	\right] \nonumber \\
	.
	\end{eqnarray}}

The difference in reflectivity between the two hemispheres has as consequence the non-coincidence of the center of mass of the satellite with the point of application of the solar radiation pressure. 
Therefore, we could have described the effect within the torque (\ref{eq_off}) analyzed in previous section (\ref{cm_diff}).  However, we preferred to maintain the expressions for two distinct torques, as done by previous authors, in order to better connect the torque to the measurable parameters, as $\Delta \rho$ for this second torque.

As we can see from Eq. (\ref{eq_asym}), also this torque is slowly changing with a period of one solar year and it is modulated by the shadow function $\nu$.

\subsection{The solution with averaged equations}
\label{av_sol}
The numerical solution of the equations (\ref{Eu_eq1} - \ref{Eu_eq3}) can be more quickly found, if averaged expressions are adopted for the torques.  This simplification does not modify the result until the satellite spins with a time scale shorter (or comparable) than the orbital period of the satellites's and of the Earth rotational period. The averaged values for the torques are calculated by integration on these longer periods.
In this case, both  the magnetic  (\ref{eq_mag})  and the gravitational (\ref{gravity}) torques can be averaged since these torques vary with time scales that are related with the pulsation of the satellite mean motion $n$ and of the Earth's angular speed $\omega_{\oplus}$, as well as with their combinations. 

\subsubsection{Averaged Magnetic Torque}

The averaged magnetic torque can be easily calculated starting from expression (\ref{eq_mag}), and integrating each one of the harmonic components over its characteristic period. Therefore, in this limit, the variables introduced in (\ref{AD}) become: $D'=D''=0$, $A'_i= \alpha'$ and  $A''_i= \alpha''$. 
For the magnetic (averaged) torque expression we obtain:

{\small 	\begin{eqnarray}
	\left<{\bf M_{mag}} \right>&=& -V <\left| {\bf B}\right|^2> \alpha''\left(\omega_s\right) \frac{ {\bf \omega_s}}{ \left| {\bf \omega_s} \right|} + V \frac{\alpha'(0) -\alpha'(\omega_s)}{\left| {\bf \omega_s} \right|^2}  \, \left<\left( {\bf B} \cdot  {\bf \omega_s} \right) \left(  {\bf \omega_s}\times  {\bf B} \right) \right> +  V \frac{\alpha''(\omega_s)} { \left| {\bf \omega_s} \right|}   \left<\left( {\bf B} \cdot  {\bf \omega_s} \right)   {\bf B} \right>= \nonumber\\
	&=&-V  \alpha''\left(\omega_s\right) \frac{ {\bf \omega_s}}{ \left| {\bf \omega_s} \right|} \left(<\left| {\bf B}\right|^2>-\left< {\bf B} \,  {\bf B}^T \right> \right)+ V \frac{\alpha'(0) -\alpha'(\omega_s)}{\left| {\bf \omega_s} \right|^2}  \,  {\bf \omega_s}\times \left(\left< {\bf B} \,  {\bf B}^T\right>  {\bf \omega_s} \right), 
	\label{tor_med} 
	\end{eqnarray}}

where we used the two relations:

\begin{equation}
\left<\left( {\bf B} \cdot  {\bf \omega_s} \right) \left(  {\bf \omega_s} \times  {\bf B} \right) \right> = {\bf \omega_s}\times \left(\left< {\bf B} \,  {\bf B}^T\right>  {\bf \omega_s} \right),
\label{med_mag1}
\end{equation}

\begin{equation}
\left<\left( {\bf B} \cdot  {\bf \omega_s} \right)   {\bf B}  \right> =\left< {\bf B} \,  {\bf B}^T \right>  {\bf \omega_s}.
\label{med_mag2}
\end{equation}    

By averaging on the nine frequency characteristic of the magnetic field, see Eq. (\ref{freq_mag}), we get the different terms to be inserted in (\ref{tor_med}), that is:

	\begin{eqnarray}
	\left\langle B_i B_j \right\rangle \left(\frac{a^6}{d^2}\right) = \alpha_{i,j} &=& \frac{1}{8}\cdot \left[9 \sin^2I+\frac{1}{2}\sin^2\theta_p( 20-27~\sin^2I)\right]\delta_{i,j}\\ 
	&&-\frac{1}{8}\left(1-3\cos^2\theta_p \right)E_iE_j\nonumber \\ 
	&&+\frac{3}{8}\cos I \left(1-3\cos^2\theta_p\right)\left(E_i\hat{n}_j+E_j\hat{n}_i \right) \nonumber \\
	&&-\frac{3}{8} \hat{n}_i \hat{n}_j\left[3\left(1-3\cos^2I\right)-\frac{1}{2}\sin^2\theta_p(5-27\cos^2I)\right], \nonumber 
	\label{alfar}
	\end{eqnarray}
	
	\begin{eqnarray}
	\left\langle B^2 \delta_{i,j} - B_i B_j \right\rangle \left(\frac{a^6}{d^2}\right)&=&     \beta_{i,j}=\left\{1+\frac{1}{8}\cdot\left[3 \sin^2I+\frac{1}{2}\sin^2\theta_p( 4-9~\sin^2I)\right]\right\}\delta_{i,j} \nonumber \\
	&&+\frac{1}{8}\left(1-3\cos^2\theta_p \right))E_iE_j  \\ 
	&&-\frac{3}{8} \cos I \left(1-3\cos^2\theta_p \right) \left(E_i \hat{n}_j+E_j \hat{n}_i \right) \nonumber \\
	&&+\frac{3}{8} \hat{n}_i \hat{n}_j\left[3\left(1-3\cos^2I\right)-\frac{1}{2}\sin^2\theta_p(5-27\cos^2I)\right], \nonumber
	\label{beta}
	\end{eqnarray}
	
	\begin{eqnarray}
	<\left| {\bf B}\right|^2>=<\sum^{8}_{i=0} \left| {\bf B_i }\right|^2>&=&\sum_1^3{\alpha_{i,i}}=-\frac{1}{4}\left[ 10-6 \cos^2I-\sin^2\theta_p (3-9\cos^2I)\right], \nonumber \\
	\end{eqnarray}

where
$ {\bf E}=[0,0,1]$ is the unit vector along the Earth rotation axis, $\delta_{i,j}$ is kronecker's delta and $ {\bf \hat{n}}$ is the  mean motion unit vector  normal to the orbital plane of the satellite.

We introduced $\alpha_{i,j}$ and $\beta_{i,j}$ to use  the same notation of expressions (7a) and (7b) of \cite{1996JGR...10117861F}. 
The first line of Eq. (\ref{alfar}) differs from the analogous expression (7a) and also the last line of Eq. (\ref{beta}) differs from the correspondent line of (7b). 
We are confident in our calculations, that follow a very different path with respect to those of  \cite{1996JGR...10117861F}, and we believe that the expressions (7a) and (7b) by \cite{1996JGR...10117861F} contain some small errors, as they do not satisfy some obvious conditions. In fact, $\beta_{i,j}+\alpha_{i,j }$ has to be diagonal and equal to $\sum_1^3{\alpha_{i,i}} $. Conversely, these conditions are satisfied by our expressions.

\subsubsection{Averaged Gravitational Torque}
\label{av_grav}
The gravitational torque (\ref{gravity}) has one constant component and one varying at a frequency twice of the satellite mean motion $n$. Integrating on the corresponding period is therefore possible to calculate the averaged torque, we obtain:

	\begin{eqnarray}
	\label{tor_grav_med}
	\left< {\bf M_{grav}}\right>&=&\frac{3}{32} n^2
	\left| 
	\begin{array}{l}
	(I_y-I_z)( M_1\cos( \psi) - M_2\sin( \psi))\\
	(I_z-I_x)(M_1\sin( \psi) + M_2\cos( \psi))\\
	(I_x-I_y)(-M_3\sin(2 \psi) - M_4\cos(2 \psi))
	\end{array}
	\right|,\nonumber\\
	\end{eqnarray}
	where	
	{\small \begin{eqnarray}
		\nonumber \\
	M_1&=&2 \left\{3 \cos(2 I)+\cos(2 \phi-2 \Omega) \left[cos(2 I)-1\right]+1\right\} \sin(2 \theta)-8 \sin(2 I) \cos(\phi-\Omega) \cos(2 \theta) \nonumber\\
	M_2&=&4 \left[cos(2 I)-1\right] \sin(2 \phi-2 \Omega) \sin(\theta)-8 \sin(2 I) \sin(\phi-\Omega) \cos(\theta)\nonumber\\
	M_3&=&\left[1+3 cos(2 I)\right] \left[cos(2 \theta)-1\right] +cos(2 \phi-2 \Omega) \left[cos(2 I)-1\right] \left[cos(2 \theta)+3\right]+4 sin(2 I) cos(\phi-\Omega) sin(2 \theta)\nonumber\\
	M_4&=&4 \sin(2 \phi-2 \Omega) \left[\cos(2 I)-1\right] \cos(\theta)+8  \sin(\phi-\Omega) \sin(2 I) \sin(\theta).
	\end{eqnarray}}

\section{Solutions of the equations for the torques}
\label{solut}
The equations (\ref{Eu_eq1}-\ref{Eu_eq3}) cannot be integrated analytically. We built a code based on MATLAB routines in order to solve for the differential equations. 
We used as independent variables the three Euler angles $\theta$, $\phi$,  $\psi$ and their time derivatives  $\dot{\theta}$,  $\dot{\phi}$ and $\dot{\psi}$, in such a way to transform the solution into that of a system of six first-order differential equations.

\subsection{Physical quantities and initial conditions}
\label{param}
A very critical point in solving the equations is in choosing the most likely initial conditions and in defining, with the highest possible precision, the parameters that appear in them. 

The moments of inertia $I_x, I_y,I_z$ of the two LAGEOS were not directly measured before their launch,  but were estimated with a careful analysis by \cite{2016AdSpR..57.1928V}. Unfortunately, we could not found any document containing the values for the moments of inertia of LARES.  In this regard, \cite{6575147} computed the moment of inertia of LARES in the hypothesis that the satellite is an homogeneous and symmetrical sphere (with no oblateness). They obtained $I = 5.125\;kg \cdot m^2$. 
With an approach a similar to the one adopted for the two LAGEOS in \cite{2016AdSpR..57.1928V}, we built also for LARES a 3D-model and  we calculated the values reported in Table \ref{tab:mass_moment}. The reported error is calculated considering that the structure of LARES was designed with the possibility to add small masses to tune the moments of inertia. In the simulations we opted for a  very small oblateness, but a spherical symmetry is nevertheless possible within the errors. 

\begin{table}[h!]
	\centering
	\caption{Mechanical parameters used in the equations: moments of inertia ${\bf I}$, ray $R$ and offset {\bf h} of the satellites.}
	\label{tab:mass_moment}
	\begin{tabular}{|c|c|c|c|}
		\hline
		&LAGEOS&LAGEOS II&LARES\\
		\hline
		$I_x\; [kg \cdot m^2]$& $10.96\pm0.03$  & $11.00\pm0.03$ & $4.76\pm0.03$ \\
		$I_y \;[kg \cdot m^2]$ &  $10.96\pm0.03$ &  $11.00\pm0.03$ &  $4.76\pm0.03$  \\
		$I_z \;[kg \cdot m^2]$&  $11.42\pm0.03$  &  $11.45\pm0.03$ & $4.77\pm0.03$  \\
		$R \;[cm]$ &30 &30 &18.2\\ 
		$h_x \;[cm]$ &0.000 &0.000 &0.000 \\ 
		$h_y \;[cm]$ &0.000 &0.000 &0.000 \\ 
		$h_z \;[cm]$ &0.040 &0.055 &0.000\\
		\hline
	\end{tabular}
\end{table}

In Table \ref{tab:em} we report the values of the electromagnetic quantities we used in our analyses and simulations. We started from the range of possible changes of the values reported in \cite{2007Andres}, modifying them within their errors with the aim to minimize the residuals between the predictions of our model and the available observations.

\begin{table}[h!]
	\centering
	\caption{Electromechanical parameters used in the equations: dimensionless magnetic factors $\beta'$ and $\beta''$, electrical conductivity $\sigma$ and the relative magnetic permeability $\mu_r$.}
	\label{tab:em}
	\begin{tabular}{|c|c|c|c|}
		\hline
		&LAGEOS&LAGEOS II&LARES\\
		\hline
		$\beta'$ & $<10^{-2}$& $<10^{-2}$  &1\\
		$\beta''$ & 0.22&0.23  &1\\
		$\sigma [s]$  & $2.37 \cdot 10^{17}$ &$2.38 \cdot 10^{17}$&$5.1 \cdot 10^{16}$\\
		$\mu_r-1$ & $2.2\cdot10^{-5}$ & $2.2\cdot10^{-5}$ &$3.3\cdot10^{-7}$\\
		\hline
	\end{tabular}
\end{table}
In Table \ref{tab:optical}, the overall optical properties of the surface of each satellite are  reported. These values are the ones that we found to be the most likely in the literature, and that we confirmed with an independent analysis.

\begin{table}[h!]
	\centering
	\caption{Optical parameters used in the equations: radiation coefficient $C_R$ and reflectivity difference between the hemispheres $\Delta \rho$ of the satellites.}
	\label{tab:optical}
	\begin{tabular}{|c|c|c|c|}
		\hline
		&LAGEOS&LAGEOS II&LARES\\
		\hline
		$C_R$ & 1.13&1.12&1.07 \\
		$\Delta \rho$ &0.013&0.012&0\\
		\hline
	\end{tabular}
\end{table}

The data of Table \ref{tab:spin_par} relative to the initial conditions for the spin vector of the two LAGEOS satellites are taken from \cite{2007Andres}, and modified within their variability interval to fit the experimental data.
Finally, in Table \ref{tab:orb_par}, the values of the keplerian parameters of the orbit of the three satellites are shown. These values were calculated averaging  the output values from a precise orbit determination (POD) made using the GEODYN II software (\cite{1990Putney,1998pavlis}).

\begin{table}[h!]
	\centering
	\caption{Spin initial conditions: reference epoch in Modified Julian Date (MJD), rotational period $P_s$, right ascension RA  and declination dec.}
	\label{tab:spin_par}
	\begin{tabular}{|c|c|c|c|}
		\hline
		&LAGEOS&LAGEOS II&LARES\\
		\hline
		Epoch [MJD] & 42913.5 &48918&55970\\
		$P_s$ [s]& 0.48 &0.81&11.8 \\
		RA [degree]& 150&230&186.5 \\
		dec [degree]& -68&-81.8&-73 \\
		\hline
	\end{tabular}
\end{table}

\begin{table*}
	\centering
	\caption{Orbital parameters used in the equations}
	\label{tab:orb_par}
	\begin{tabular}{|c|c|c|c|}
		\hline
		&LAGEOS&LAGEOS II&LARES\\
		\hline
		Day of reference [MJD] &48989 &49003 &55975 \\
		Semimajor axis $a$ [cm]& $1.2270 \cdot 10^{9}$ &$1.2162 \cdot 10^{9}$&$7.82035\cdot 10^{8}$\\
		Eccentricity $e$ & 0.004  & 0.014 & 0.001\\
		Inclination $I$ [degree]& 109.84  &52.66&69.49\\
		Ascending node longitude $\Omega$ [degree]& 313.72&60.62&236.4\\
		Ascending node longitude rate $\dot{\Omega}$ [degree/day]& 0.34 &-0.63&$-1.71$\\
		Argument of pericenter  $\omega$ [degree]& 39.90 &251.82&296.055\\
		Argument of pericenter  rate $\dot{\omega}$ [degree/day] & -0.21  &0.44&-0.95\\
		mean anomaly $M_0$ [degree] & 79.51&103.36&63.933\\
		\hline
	\end{tabular}
\end{table*}

\subsection{Numerical solution and experimental data}\label{num_sol}
In this section we present our results for the time evolution of the spin of the two LAGEOS and LARES satellites. This evolution has been calculated using the equations of motion previously introduced, namely Eqs. (\ref{Eu_eq1}), (\ref{Eu_eq2}) and (\ref{Eu_eq3}).

The expressions that we used for the torques were of two types: the more general one, that we have discussed in section \ref{Gen_sol}, and the averaged one (see section \ref{av_sol}). This last solution differs from the general one for the magnetic and gravitational torques. For the other two torques that we considered, that vary with the periodicity of one year and are locally modulated by the Earth's shadow function, we adopted the general expressions (\ref{eq_off}) and (\ref{eq_asym}). 

\begin{figure}[htb!]
	\centering
	\includegraphics[width=12cm]{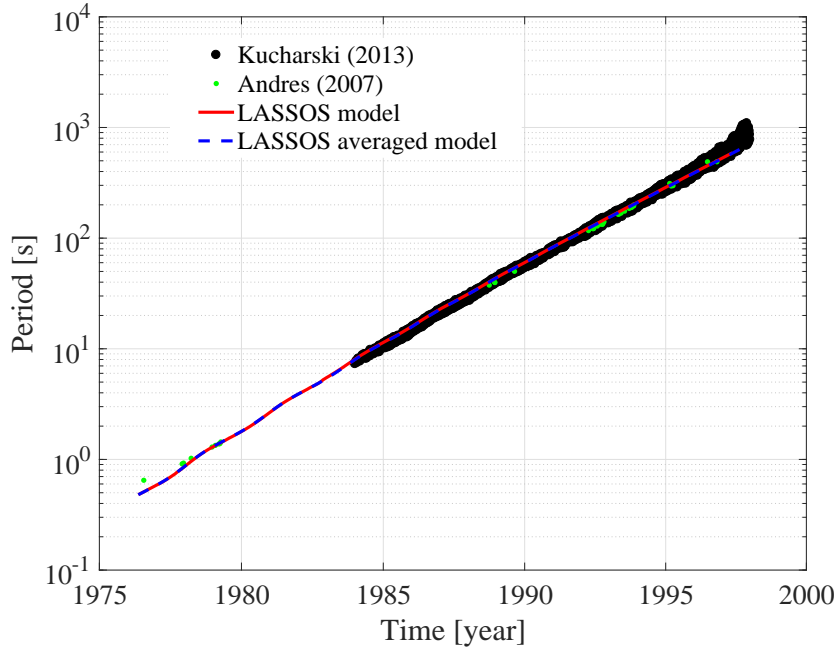}
	\caption{Spin period of LAGEOS. The continuous red line is the result of the numerical integration using our new model LASSOS, which is based on general expressions for the torques. The dashed blue line represents the solution obtained using averaged torques. The black points are measurements got from Fig.3 of \cite{2013AdSpR..52.1332K}. The red crosses are measurements reported by \cite{2007Andres}.
		\label{fig:LG1_pe}}
\end{figure}

\begin{figure}[h!]
	\centering
	\includegraphics[width=15cm]{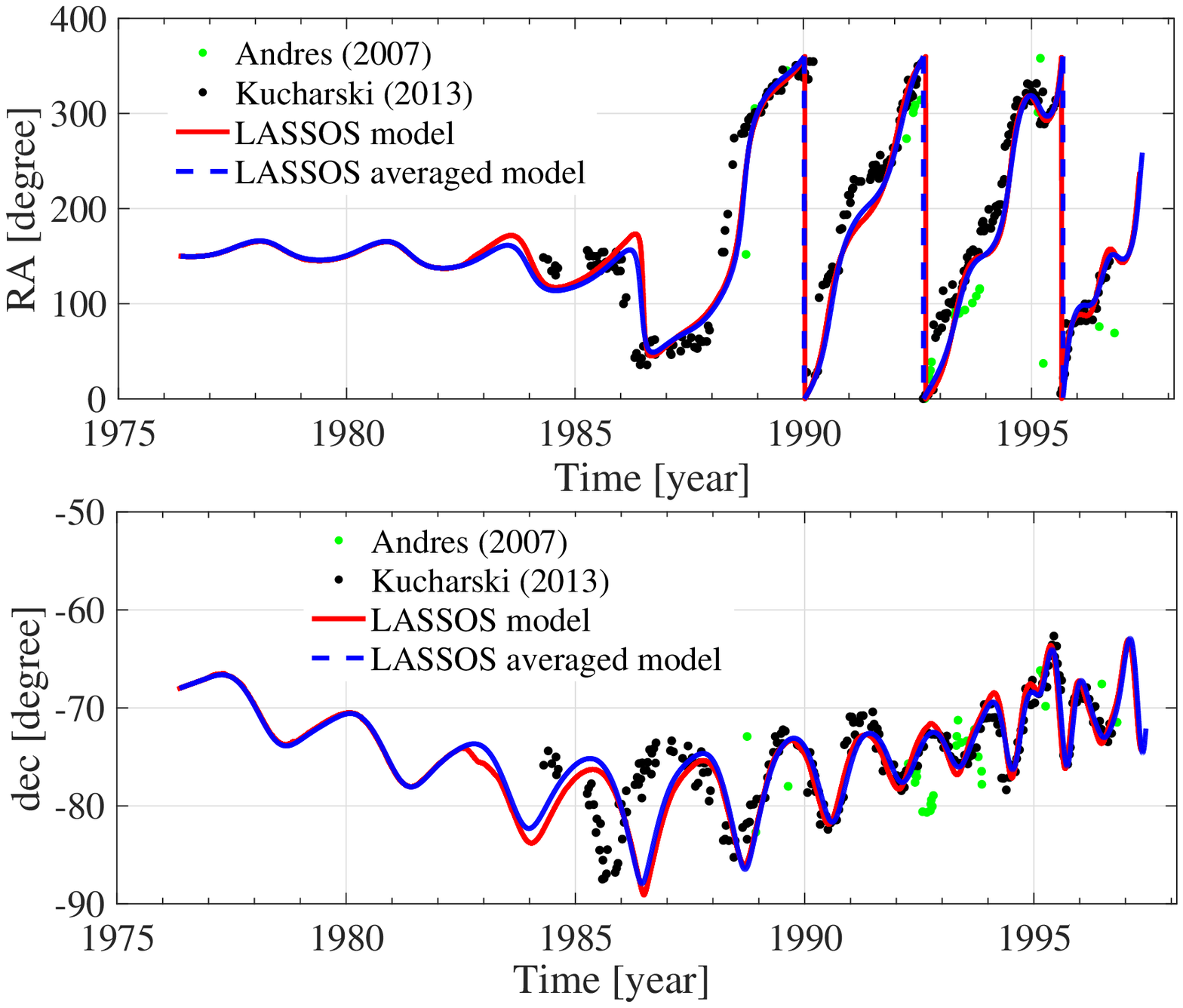}
	\caption{Right ascension (top) and declination (bottom) of LAGEOS. The continuous red line is the result of the numerical integration using our new model LASSOS, which is based on general expressions for the torques. The dashed blue line represents the solution obtained using averaged torques. The black points are measurements reported by \cite{2013AdSpR..52.1332K}. The red crosses are measurements reported by \cite{2007Andres}.}
	\label{fig:LG1_rd}
\end{figure}

\begin{figure}[h!]
	\centering
	\includegraphics[width=15cm]{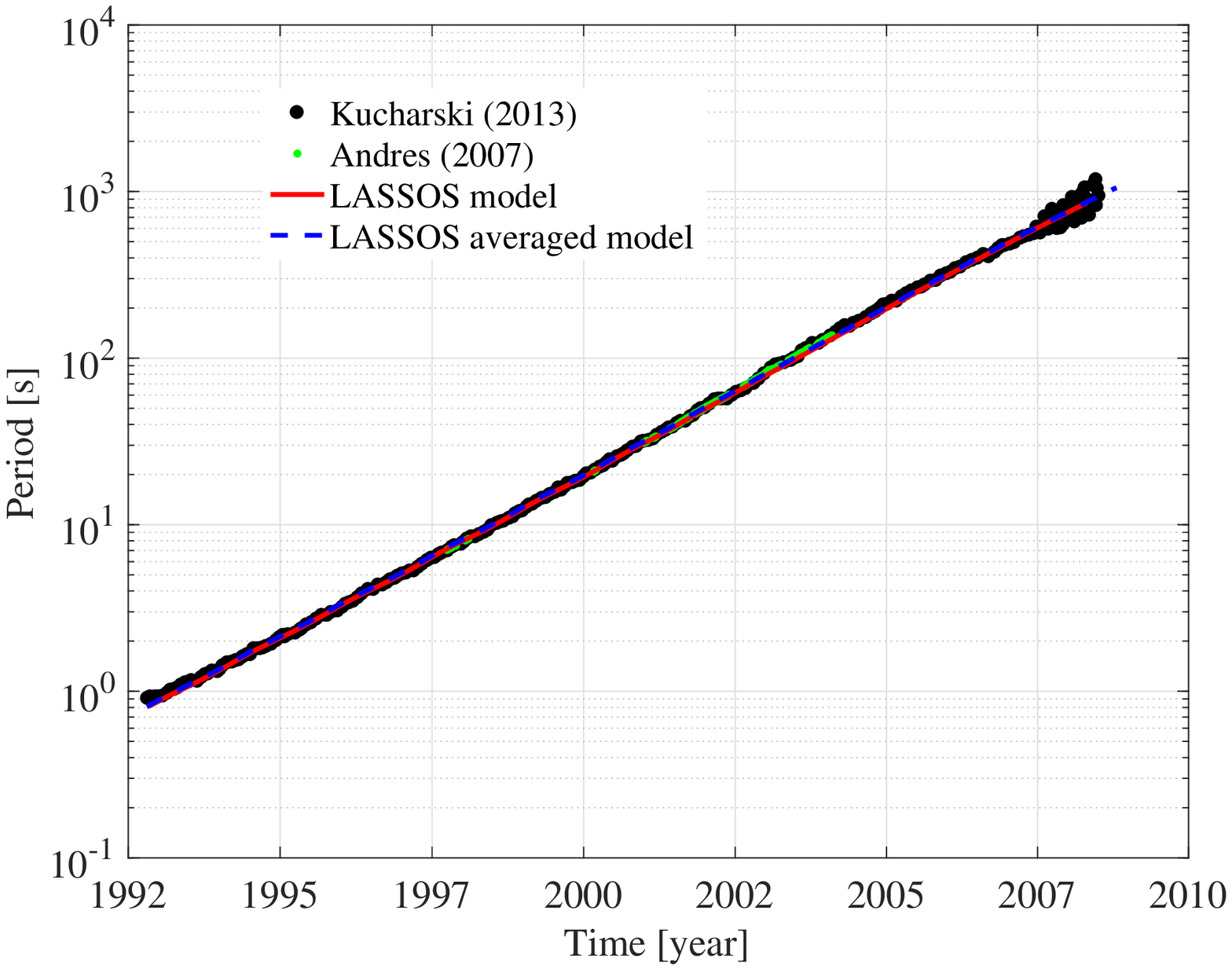}
	\caption{Spin period of LAGEOS II. The continuous red line is the result of the numerical integration using our new model LASSOS, which is based on general expressions for the torques. The dashed blue line represents the solution obtained using averaged torques. The black points are measurements reported by \cite{2013AdSpR..52.1332K}. The red crosses are measurements reported by \cite{2007Andres}.}
	\label{fig:LG2_pe}
\end{figure}

\begin{figure}[h!]
	\centering
	\includegraphics[width=15cm]{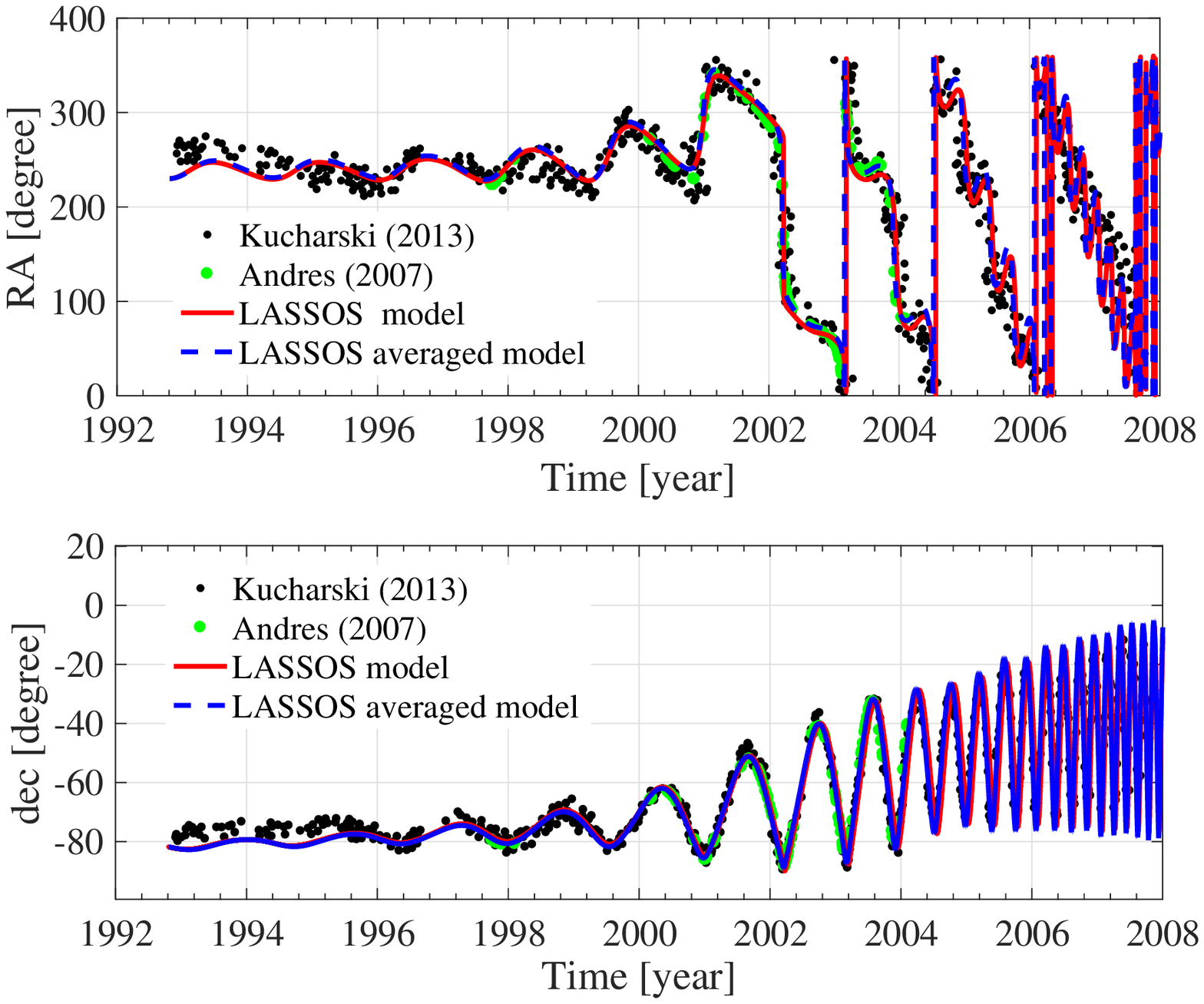}
	\caption{Right ascension (top) and declination (bottom) of LAGEOS II. The continuous red line is the result of the numerical integration using our new model LASSOS, which is based on general expressions for the torques. The dashed blue line represents the solution obtained using averaged torques. The black points are measurements reported by \cite{2013AdSpR..52.1332K}. The red crosses are measurements reported by \cite{2007Andres}.}
	\label{fig:LG2_rd}
\end{figure}

\begin{figure}[h!]
	\centering
	\includegraphics[width=15cm]{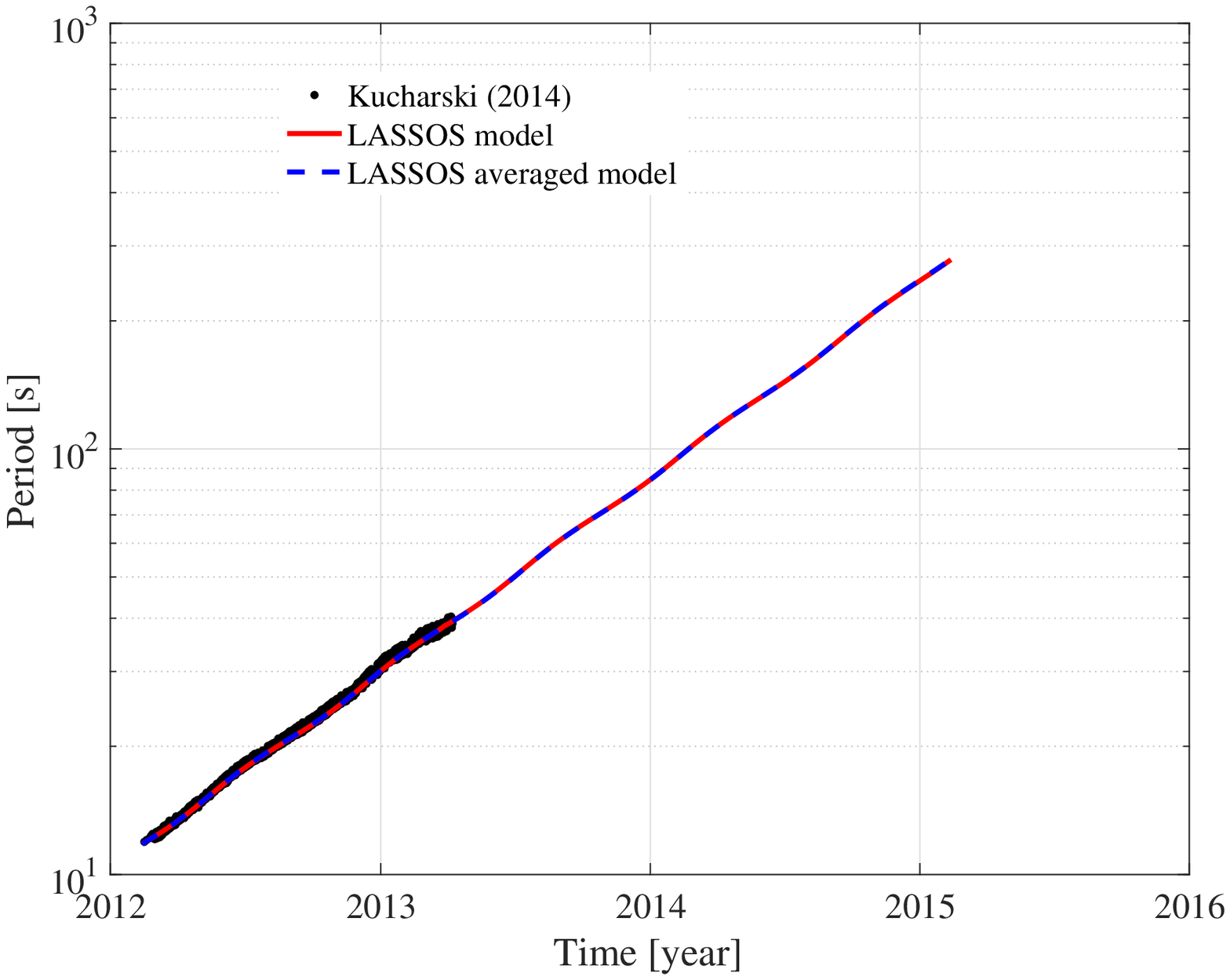}
	\caption{Spin period of LARES. The continuous red line is the result of the numerical integration using our new model LASSOS, which is based on general expressions for the torques. The dashed blue line represents the solution obtained using averaged torques.  The black points are measurements from \cite{6575147}. }
	\label{fig:LR1_pe}
\end{figure}

\begin{figure}[h!]
	\centering
	\includegraphics[width=15cm]{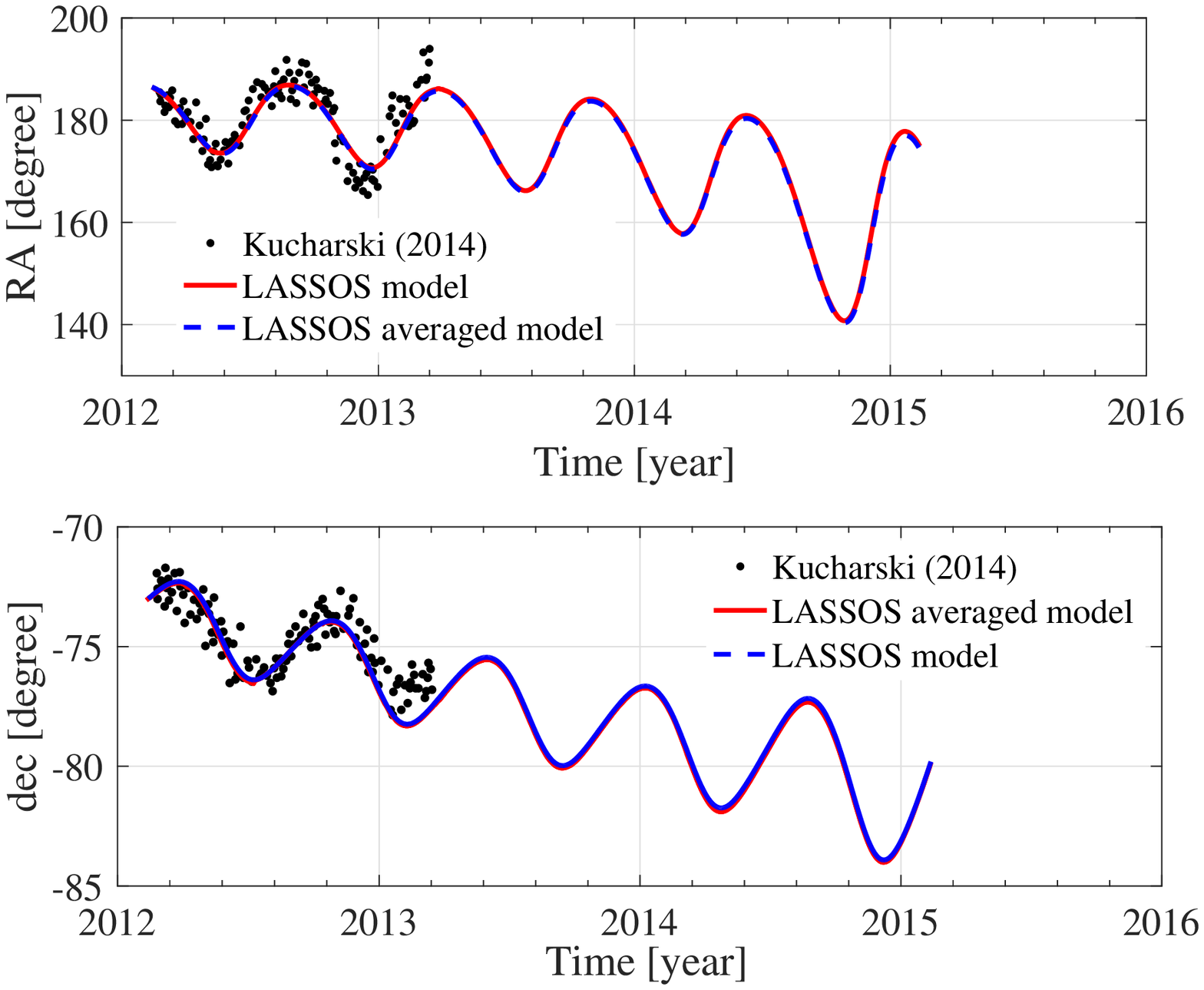}
	\caption{Right ascension (top) and declination (bottom) of LARES. The continuous red line is the result of the numerical integration using our new model LASSOS, which is based on general expressions for the torques. The dashed blue line represents the solution obtained using averaged torques.  The black points are measurements from \cite{6575147}.}
	\label{fig:LR1_rd}
\end{figure}

The parameters and the initial conditions we used are those reported in Tables 
\ref{tab:mass_moment}--\ref{tab:orb_par}. 
For each satellite it has been calculated and drawn the period of rotation and the orientation (right ascension and declination) of the spin as function of time. 
In Figures (\ref{fig:LG1_pe}) and (\ref{fig:LG1_rd}) we show the results for the older LAGEOS, while in  Figures (\ref{fig:LG2_pe}), (\ref{fig:LG2_rd}) and in Figures (\ref{fig:LR1_pe}), (\ref{fig:LR1_rd}) we show, respectively, the results in the case of LAGEOS II and for the newly LARES.  



In all these plots the results for both the general model (red continuous line) and the averaged model (blue dashed line) are shown. These results are compared with all the available measurements. 
All these comparisons validate very well our models. 
As specified at the beginning of section \ref{our_sol}, we refer to our new general model as to LASSOS, that is the acronym of LArase Satellites Spin mOdel Solutions.

The bulk of the experimental measurements are of two kind for the two LAGEOS: those obtained using Sun's light reflected by the CCR mirrors (see \cite{1997PhDT........14A}), and those obtained observing the modulation of the laser light reflected back by the satellite's CCR (see \cite{GRL:GRL14288}). The last measurements are indeed available for all three satellites (see \cite{2012AdSpR..50.1473K,2013AdSpR..52.1332K}).
The latter method has increased a lot in efficiency after the introduction of high frequency (kHz) repetition laser in several SLR stations, the first of these stations has been in 2003 Graz.

In the comparison between models and observations, a disagreement is present for LAGEOS's declination in the years 1983-1987. The reason for this disagreement with the data analyzed by \cite{2013AdSpR..52.1332K} could come from the still low quality of the laser used by the ILRS stations in that years. 
Another disagreement is present in the case of LAGEOS in the  measurements after 1996, and for LAGEOS II after 2008. These periods correspond to the decrease of the spin frequency of the satellites, close to the limit of the possible measurements with kHz laser. 
We notice that the same disagreements between experimental measurements and model are present when the LOSSAM model is applied, see Figure 1  of \cite{2013AdSpR..52.1332K}.

\begin{figure}[h!]
	\centering
	\includegraphics[width=15cm]{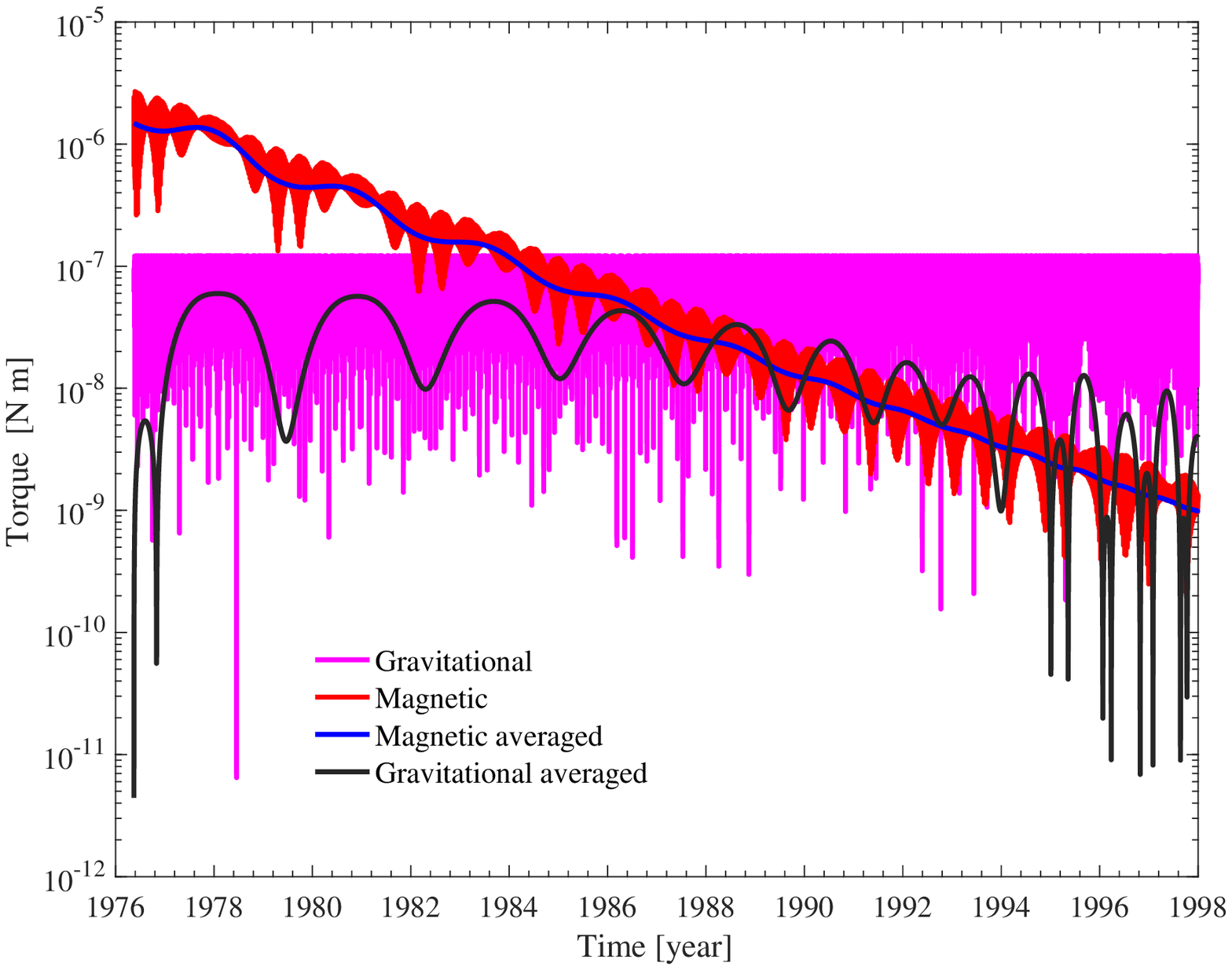}
	\caption{Absolute value (modulus) of the gravity gradient and magnetic torques acting on LAGEOS during the first 20 years of its life in J2000 reference frame. }
	\label{fig:mod_torque_GM}
\end{figure}

\begin{figure}[h!]
	\centering
	\includegraphics[width=15cm]{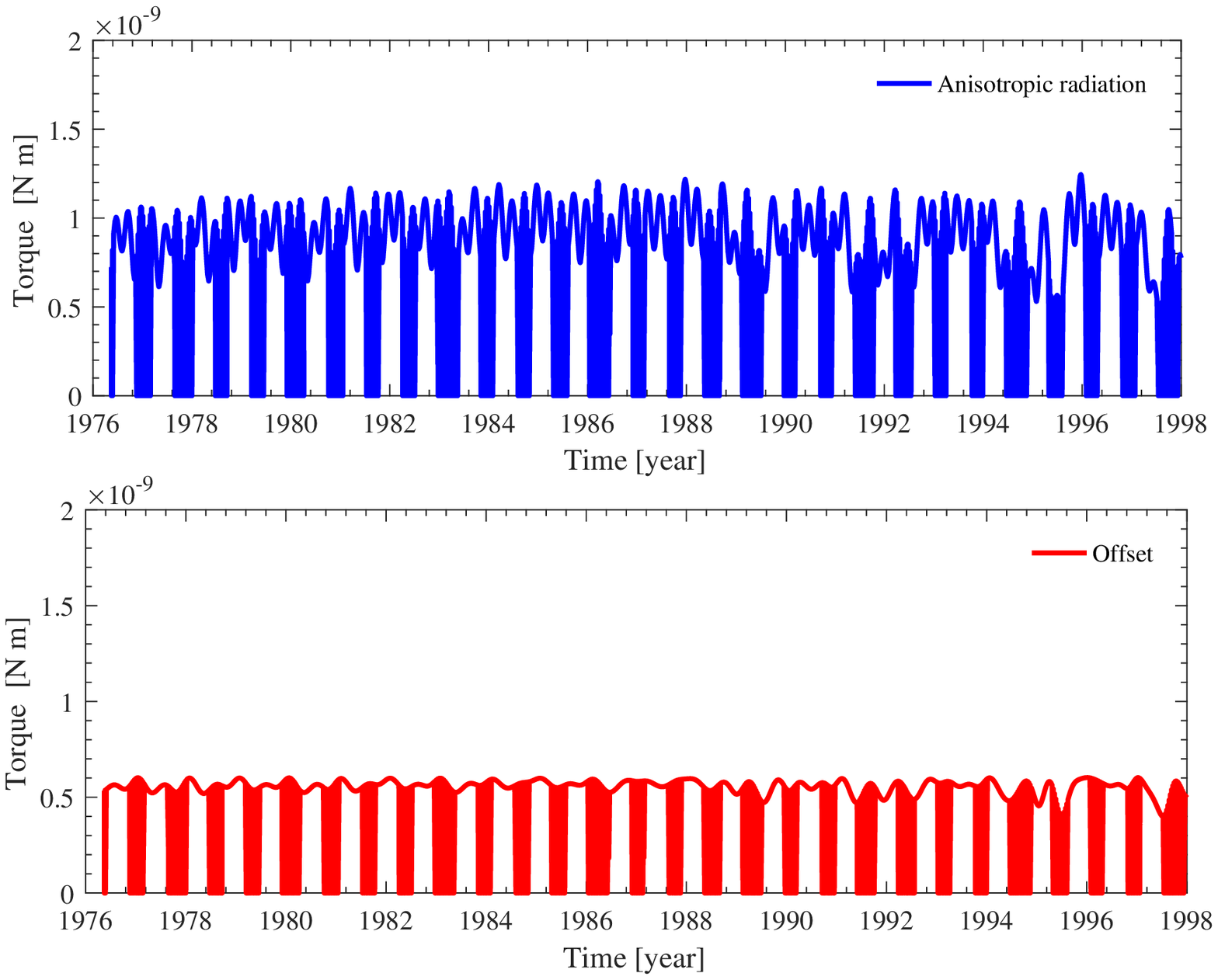}
	\caption{Absolute value (modulus) of the two torques due to radiation pressure acting on LAGEOS during the first 20 years of its life in J2000 reference frame.}
	\label{fig:mod_torque_o_a}
\end{figure}

Using our model LASSOS we have also computed the time evolution of the absolute value and of the components of the different torques acting on the satellites in the J2000  reference frame.  
For instance, in the case of LAGEOS in Figure (\ref{fig:mod_torque_GM}) we show the modulus of the torques due to the Earth's gravity gradients and magnetic field, both when they are averaged and when they are not averaged, while in Figure (\ref{fig:mod_torque_o_a}) the values of the two torques due to radiation pressure are shown for the same period. The amplitude of these torques is modulated by the eclipses, that bring these torques to zero value.
During the first period the dominating torque is the magnetic one, later it is overcome by the  torque from gravity gradients. In the last years, the effect of the torques that arise from  radiation pressure becomes comparable with that of the other two torques.


\begin{figure}[h!]
	\centering
	\includegraphics[width=15cm]{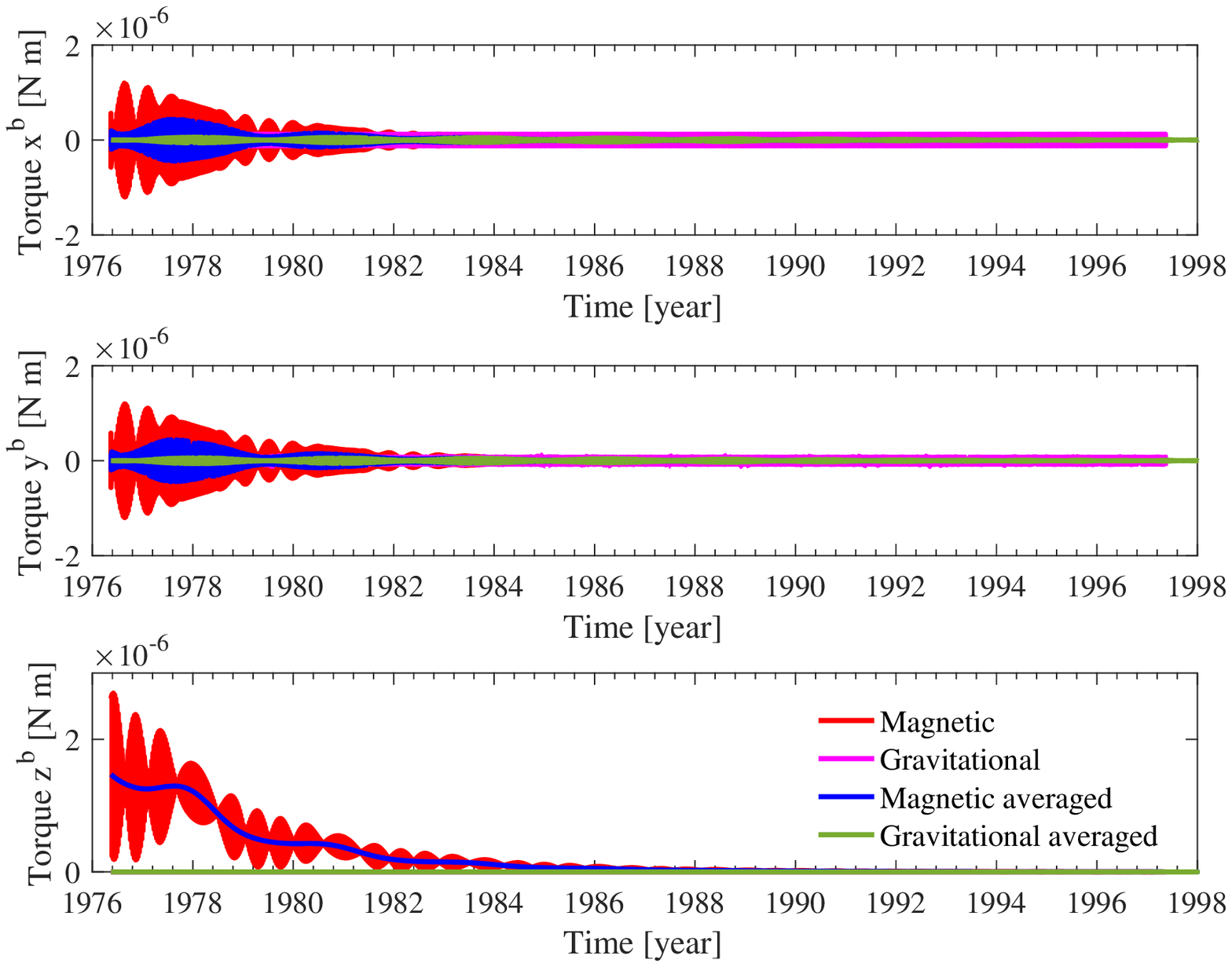}
	\caption{The three components of the gravity gradients and magnetic torques acting on LAGEOS during the first 20 years of its life in the body reference frame.}
	\label{fig:torque_GM}
\end{figure}

\begin{figure}[h!]
	\centering
	\includegraphics[width=15cm]{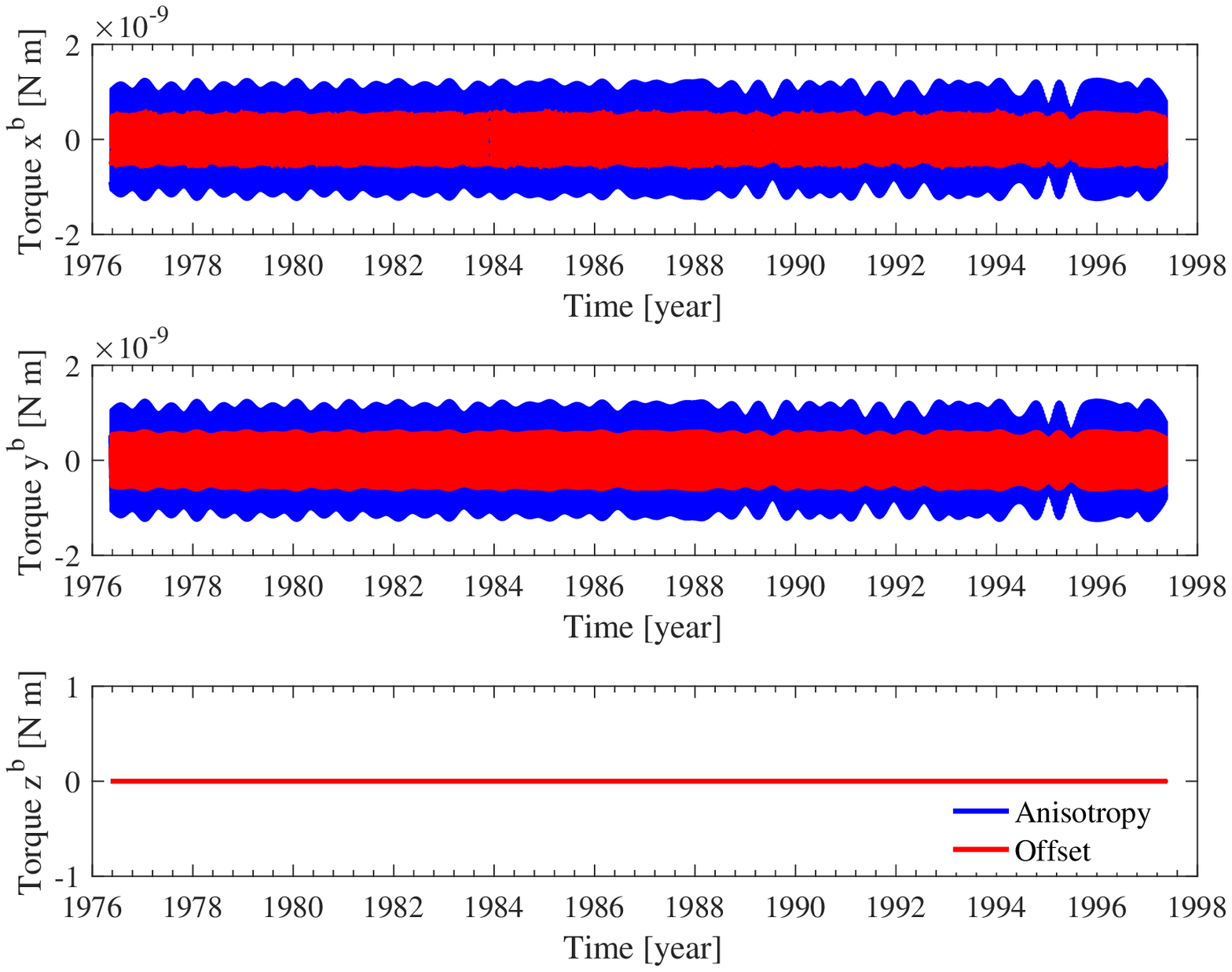}
	\caption{The three components of the radiation pressure torques acting on LAGEOS during the first 20 years of its life in the body reference frame.}
	\label{fig:torque_o_a}
\end{figure}

Finally, in Figure (\ref{fig:torque_GM}) and (\ref{fig:torque_o_a}),  the three components of the considered torques on LAGEOS during the first 20 years of its life, and in the body reference frame, are shown.
The torque due to the Earth magnetic field is the only one acting on the $ {\bf \hat{z}^b}$ axis with a despinning effect. If the geometric center, where the radiation pressure force is applied,  has coordinates $h_x$ or $h_y$ different from zero, a further component of the torque along the  $ {\bf \hat{z}^b}$ axis will appear; in this case the torque could also have an effect in increasing the spin rate of the satellite.

Similar considerations to those obtained in the case of LAGEOS are also valid for LAGEOS II and LARES.

\section{Discussion about the results}\label{dis}

In this work we have faced the problem  of the knowledge of the spin evolution of LAGEOS-like satellites. Our first goal in this field was to significantly improve the  time evolution of the spin vector of the passive LAGEOS, LAGEOS II and LARES satellites.
The knowledge of the right spin evolution for these satellites, both in rate and orientation, is of fundamental importance in order to correctly model the thermal thrust forces acting on their surface, as well as the asymmetric reflectivity observed in the case of the two LAGEOS.
The ability to manage these non-gravitational perturbations through reliable models is, consequently, of utmost importance. This not only for studies focused on measurements of fundamental physics by means of the analysis of the orbit of these satellites, but also for the potential benefits in the fields of space geodesy and geophysics.
 
For instance, in this regard, the ILRS determinations of the tracking station coordinates and of the Earth Orientation Parameters (EOP), represent the starting point for the determination of the Earth's geocenter and the definition of the Earth's International Terrestrial Frame (ITRF).
These, and other geophysical products, require a careful orbit determination of the two LAGEOS and (now) LARES (and other) satellites. Indeed, a POD should be based on the continuous refinement of the perturbations included in the dynamical model of each satellite.

In Section \ref{our_sol} we have introduced the main torques acting on the two LAGEOS and LARES satellites. Of course, the four effects there considered are not the only physical effects able to produce a torque on a spacecraft in orbit around our planet.
However, these torques are the most important to take into account at the current level of the orbit modelling of the LAGEOS and LARES satellites and also of the their corresponding POD, that depends not only by the models implemented in the software but also from the tracking observations and the Reference Frames accuracy. In fact, the other possible torques that we can take into consideration are negligible, at least by two or three order of magnitude, with respect to the torque due to the asymmetric reflectivity and the one due to the offset between the satellite center of mass and the center of pressure (i.e. the geometric center), that is  the current smallest torques that have a maximum magnitude of about $10^{-9} N m$, see Figure \ref{fig:mod_torque_o_a}.

For these reasons we have not performed a complete analysis to model torques that $\mathit{a~priori}$ should be orders of magnitude smaller than those considered, but we simply estimated the order of magnitude of the effects not included in our model. In particular, we evaluated the order of magnitude of five additional torques originated by:
\begin{itemize}
	\item Neutral drag 
	\item Charged drag 
	\item Earth Yarkovsky 
	\item Yarkovsky-Schach  
	\item Inner eddy currents
\end{itemize}
\begin{table}[h!]
		\centering
		\caption{Additional torques and their order of magnitude in the case of the two LAGEOS satellites.}
		\label{tab:torques_add}
		\begin{tabular}{|c|c|c|}
			\hline
			Perturbing effect & Acceleration $[m/s^2]$ & Torque [N m]\\
			\hline
			Neutral drag & $3\times 10^{-13}$ & $1\times 10^{-13}$ \\
			Charged drag & $5\times 10^{-12}$ & $2\times 10^{-12}$ \\
			Thermal drag & $7\times 10^{-12}$ & $3\times 10^{-12}$ \\
			Yarkovsky-Schach & $1\times 10^{-10}$ & $4\times 10^{-11}$ \\
			Inner eddy currents & $4\times 10^{-13}$ & $3\times 10^{-11}$ \\
			\hline
		\end{tabular}
\end{table}
We show in Table \ref{tab:torques_add} the order of magnitude of the acceleration and of the corresponding torque produced by these physical effects in the case of LAGEOS. These orders of magnitude are still valid in the case of LAGEOS II. The first four torques are in principle null for a satellite with a perfectly spherical in shape mass distribution. Deviations from this distribution are due to an offset between the geometrical center of the satellite and its center of mass. In the estimation of the torques that arise from these effects a very conservative value of 1 mm for the offset of the three satellites was adopted. The margin adopted in the estimation can be evaluated comparing the value of the offset of  1 mm with those used in the current work that were estimated on the basis of our fit of the available spin observations of the satellites, see previous Table \ref{tab:mass_moment}.   

The accelerations acting on the satellite  in the case of neutral drag  have been obtained from \cite{2015CQGra..32o5012L}, and from \cite{2007Andres} --- as a very pessimistic upper bound --- in the case of the drag produced by the Coulomb interaction of the charged spacecraft with the surrounding charged particles.
The thermal drag acceleration produced by the Earth-Yarkovsky effect has been obtained as an upper limit of the acceleration estimated by \cite{1988JGR....9313805R}, while in the case of the solar Yarkovsky-Schach effect the value is the one estimated by \cite{2002P&SS...50.1067L}.

These last two torques have to be considered as torques related to a thermoelastic deformation of the satellites, and are operative only if the temperature distribution across the satellite surface is effective in changing the spacecraft mass distribution.
Finally, the last effect is related with the electrodynamical force produced by the eddy currents generated by the photoelectric emission of the satellite, see also \cite{2007Andres} and \cite{2000Nakagawa}.

In Table \ref{tab:torques_add_lares} the order of magnitude of possible additional torques are shown in the case of LARES. Among the non-gravitational perturbations acting on this satellite, the effects of the neutral atmosphere drag need a special attention, since the relatively low orbit of the satellite, at an height of about 1,450 km  where the residual Earth atmosphere cannot be neglected (see \cite{2017Pardini}).

\begin{table}[h!]
	\centering
	\caption{Additional torques and their order of magnitude in the case of the LARES satellite.}
	\label{tab:torques_add_lares}
	\begin{tabular}{|c|c|c|}
		\hline
		 Perturbing effect & Acceleration \([m/s^2]\) & Torque [N m]\\
		\hline
		Neutral drag & \(1\times 10^{-11}\) & \(4\times 10^{-12}\) \\
		Other effects & \(2\times 10^{-13}\) & \(8\times 10^{-14}\) \\
		\hline
	\end{tabular}
\end{table}

The neutral drag produces on LARES an acceleration of about $1 \times 10^{-11} [m/s^2]$ and a torque, in the presence of the cited offset, of about $4 \times 10^{-12} [N m]$, a factor $\approx 40$ larger than that obtained in the case of the two LAGEOS, but still negligible with respect to the smaller torque of those considered within the LASSOS model.

The last line of Table \ref{tab:torques_add_lares} accounts for the possible cumulative contribution to the torque in the presence of an offset of about 1 mm from the other smaller perturbing effects.

\section{Conclusions and future work}\label{concl}

The work here presented falls in the LARASE research program, whose ultimate goal is to provide refined measurements of relativistic physics on the orbit of laser-ranged satellites, as the ones considered in the present paper: the two LAGEOS and LARES.
Indeed, a very important prerequisite in order to reach refined measurements of the tiny gravitational effects on the orbit of a satellite, as those predicted by Einstein's theory of General Relativity, is the inclusion of accurate models for the handling of both gravitational and non-gravitational perturbations in the data reduction of the orbit of a satellite.
In particular, the non-gravitational perturbations are responsible of very subtle effects on the orbit, really quite complex to model in a reliable fashion.
Their modeling will first impact the orbit determination of the satellites considered for the relativistic measurements of interest and, finally, will have a deep impact in the robustness of the final error budget of the measurements.

In the present work we have described the effects of the torques due to: 
\begin{itemize}
	\item{Earth's magnetic field}
	\item Earth's gravitational field	
	\item not coincidence between geometrical center and center of mass of the satellite
	\item asymmetries in the reflectivity of the satellite surface.
\end{itemize}    

The model for the spin presented in this paper, i.e. LASSOS, is, to our knowledge, the only model --- among all the models developed so far --- that has succeeded at the same time in answering to all the following aspects:

\begin{enumerate}
	\item{it implements not-averaged torques}
	\item{it can be used for any orientation of the satellite rotation axis}
	\item{it is valid for any spin rate}
	\item{it has produced results compatible with the available observations.}
\end{enumerate}

The good agreement with the experimental measurements is also consequence of the independent evaluation of the moments of inertia of the satellites (\cite{2016AdSpR..57.1928V}). 
A further innovation is the introduction of a description of the torque from Earth's magnetic field using its local values along the satellites' orbit, and not simply by means of a static average value for the magnetic field. Indeed, this represents a very important result.

A step forward of LASSOS with respect to previous models is also represented by the expression that we used for the magnetic polarizability per unit of volume $\alpha$ of the satellite.
In particular, we considered the dependence of $\alpha$ from the relative magnetic permeability $\mu_r$, that provides the residual magnetic polarizability in the case of a non-spinning satellite.
This aspect is quite important for the considered satellites. Indeed, the rotational period of LAGEOS is quite long, probably of the order of about 1 day, while for LAGEOS II it is probably of the order of its orbital period. In the case of LARES the rotational period is probably of a few thousands of seconds, but its slowing down is much faster than that of the two LAGEOS.
An average of our solution on the day and the orbital period, in such a way to reproduce the limit of the spin evolution valid in the so-called fast-spin approximation, has allowed us to point out also some errors present in previous works, which are based on simplified averaged models.

In order to further improve (if possible) the knowledge of the parameters that enter in our spin model, in the near future we plan to develop a more sophisticated tool regarding the fit procedures that we used for the comparison of the available observations (i.e. the measurements of the orientation and rate) of the spin and  the corresponding predictions by the LASSOS model.
We retain that this tool will be fundamental in order to increase the robustness for the prediction of the spin evolution in the periods not covered by measurements, and during which the spin rate is so low that only our spin model, and not one model based on averaged equations, is able to provide a reliable foresight.
Of course, the possibility of comparing the predictions of LASSOS with new observations of the spin would be very interesting. Considering the techniques applied till now for the determination of the spin of the satellites, this is not an easy task in the case of the two LAGEOS, since their very low rotational rate, but it is still possible for LARES with a spectral analysis of the range data obtained by means of kHz lasers. 

A further and very important step to be performed in the near future is to include the contribution of the predictions of the spin  provided by LASSOS in a thermal model of the satellites and compute the changes in their orbit. These changes should than be compared with the corresponding orbital residuals of the satellites obtained within a dedicated POD, in order to see how well these residuals will be reduced by modeling the thermal effects.

The development of a new thermal model for the considered satellites is one of the main targets of LARASE, and we are confident to present soon the model and the results that can be obtained applying the model to a POD.
Nevertheless, we preliminary tested the LASSOS model in this direction with a POD performed for the two LAGEOS with GEODYN in the case of the Yarkovsky thermal drag.
We obtained a slightly reduction in the RMS of the satellites range residuals over a timespan of several years: from 2.8 cm to 2.5 cm in the case of LAGEOS and from 2.5 cm to 2.2 cm for LAGEOS II.
It is important to underline that the thermal model included in Geodyn is a very old model not up-to-date, and it is valid only in the case of the fast spin approximation. Moreover, the starting epoch of the data reduction was that of LAGEOS II launch, consequently LAGEOS was not rotating fast, while LAGEOS II  was in the condition of fast rotation only during the first years of the analysis.

However, this preliminary result represents an argument in favour of the fact that we are on the right path with the LASSOS model.

\section*{Acknowledgments}
The authors acknowledge the ILRS for providing high quality laser ranging data of the two LAGEOS satellites and of LARES. Special thanks to D. Kucharski for providing the spin measurements of LARES.
We thank our colleagues of the LARASE collaboration and, in particular, R. Peron for useful discussion. This work has been in part supported by the Commissione Scientifica Nazionale II (CSNII) on astroparticle physics experiments of the italian Istituto Nazionale di Fisica Nucleare (INFN). 

\appendix
\section{The Earth magnetic field calculated along the orbit of a satellite}
If the Earth magnetic field is approximated by a dipole $ {\bf d_E}$ (with $d>0$) tilted with respect to the Earth rotational axis in such a way that $\alpha_p$ and $\theta_p$ are the longitude and colatitude of the magnetic north pole, the dipole in the J2000 reference frame evolves in time as:

\begin{eqnarray}
 {\bf d_E(t)}=d\left[\begin{array}{c} \, \cos \left(\alpha_p +\omega_\oplus\, t\right)\, \sin\!\left(\theta_p\right)\\ 
\sin\!\left(\alpha_p +\omega_\oplus\, t\right)\, \sin\!\left(\theta_p\right)\\
\cos\!\left(\theta_p\right)
\end{array}\right].
\end{eqnarray}

The satellite motion is approximated on a circular orbit with Keplerian parameters measured in J2000 frame: $M_0$, the mean anomaly at t=0; $\Omega$, the longitude of ascending node,  $I$, the orbit inclination, $\omega$, the argument of pericenter and $a$, the semimajor axis. Than the satellite motion in the J2000 frame is:

\begin{eqnarray}
 {\bf s}=a\cos\left(M_0 + n t\right)\,
\left[\begin{array}{c}
\cos \Omega\, \cos \omega - \cos I \,\sin \Omega\,\sin \omega\\
\sin \Omega\, \cos \omega + \cos I \, \cos \Omega\,\sin \omega\\
\,\sin I \,\sin \omega \\
\end{array}
\right]+ \nonumber\\
- a \sin\left(M_0 +  n t\right)\,\left[
\begin{array}{c}
\cos \Omega\,\sin \omega + \cos I\,\sin \Omega\, \cos \omega\\
\sin \Omega\,\sin \omega - \cos I\, \cos \Omega\, \cos \omega\\
- \sin I \, \cos \omega
\end{array}\right].
\end{eqnarray}

The magnetic field as function of time over the satellite orbit is therefore given by:

	\begin{eqnarray}
	 {\bf B_E}=\frac{3\,  {\bf s} ( {\bf d_E} \cdot  {\bf s})}{a^5}-\frac{ {\bf d_E}}{a^3}= \sum^{8}_{i=0}  {\bf B_i } \cos(\omega_i t+ \varphi_{i}), 
	\end{eqnarray}
	where
	\begin{eqnarray}
	\label{freq_mag_App}
	\omega_0&=&0 \nonumber\\
	\omega_1&=&\omega_2=\omega_{\oplus}-2 n  \hspace{30pt} \nonumber\\
	\omega_3&=&\omega_4=\omega_\oplus+2 n \hspace{30pt}  \\
	\omega_5&=&\omega_6=2 n \hspace{30pt}  \nonumber \\
	\omega_7&=&\omega_8=\omega_\oplus \nonumber
	\nonumber \\
	\varphi_{i}&=&\left\{
	\begin{array}{cc}
	-\frac{\pi}{2}& \mbox{for i=2,4,6,8} \\
	0& \mbox{for i=0,1,3,5,7} 
	\end{array} 
	\right.
	\end{eqnarray}
\linebreak
{\small 	
	\begin{eqnarray}
	 {\bf B_0}&=&  - \frac{3\, d \cos~\theta_p }{4 \, a^{3}}    \left[ 
	\begin{array}{c}
	-\sin(2I) \sin \Omega\\
	\sin(2I) \cos \Omega\\
	-\frac{1}{3} + \cos(2I) 
	\end{array}
	\right]
	\\
	\nonumber \\
	 {\bf B_{1}}&=&- \frac{3\, d \sin~\theta_p \left(1+\cos~I \right)}{8 \, a^{3}}      \left[
	\begin{array}{c}
	(1-\cos~I) \sin(2\omega+2M_0-\phi_p)+(1+\cos~I) \sin(2\Omega+2\omega+2M_0-\alpha_p) \\
	(1-\cos~I)\cos(2\omega+2M_0-\phi_p)-(1+\cos~I)\cos(2\Omega+2\omega+2M_0-\alpha_p)  \\
	-2 \sin~I \cdot \cos(\Omega +2\omega+2M_0-\alpha_p)  
	\end{array}
	\right] \nonumber \\
	\\
	\nonumber \\
	 {\bf B_{2}}&=&- \frac{3\, d \sin~\theta_p \left(1+\cos~I \right)}{8 \, a^{3}}     \left[
	\begin{array}{c}
	(1-\cos~I)\cos(2\omega+2M_0-\phi_p)+(1+\cos~I) \cos(2\Omega+2\omega+2M_0-\alpha_p)  \\
	-(1-\cos~I)\sin(2\omega+2M_0-\phi_p)+(1+\cos~I) \sin(2\Omega+2\omega+2M_0-\alpha_p)\\
	2~ \sin~I \cdot \sin(\Omega + 2\omega+2M_0-\alpha_p) 
	\end{array}
	\right] \nonumber \\
	\\
	\nonumber \\
	 {\bf B_{3}}&=&- \frac{3\, d \sin~\theta_p \left(1-\cos~I \right)}{8 \, a^{3}}    \left[
	\begin{array}{c}
	-(1+\cos~I)\sin(2\omega+2M_0+\phi_p)+(1-\cos~I)\sin(2\Omega-2\omega-2M_0-\alpha_p)\\
	(1+\cos~I)\cos(2\omega+2M_0+\phi_p)-(1-\cos~I)\cos(2\Omega-2\omega-2M_0-\alpha_p)\\
	2~ \sin~I \cdot \cos(\Omega-2\omega-2M_0-\alpha_p)
	\end{array}
	\right] \nonumber \\
	\\
	\nonumber \\
	 {\bf B_{4}}&=&- \frac{3\, d \sin~\theta_p \left(1-\cos~I \right)}{8 \, a^{3}}    \left[
	\begin{array}{c}
	(1+\cos~I)\cos(2\omega+2M_0+\phi_p)+(1-\cos~I)\cos(2\Omega-2\omega-2M_0-\alpha_p)\\
	(1+\cos~I)\sin(2\omega+2M_0+\phi_p)+(1-\cos~I)\sin(2\Omega-2\omega-2M_0-\alpha_p)\\
	-2 ~\sin~I \cdot \sin(\Omega - 2\omega-2M_0-\alpha_p)
	\end{array}
	\right] \nonumber \\
	\\
	 {\bf B_{5}}&=& - \frac{3\, d \cos~\theta_p \sin~I }{2 \, a^{3}}   \left[
	\begin{array}{c}
	\cos \Omega \cos(2\omega+2M_0)- \cos~I \sin(\Omega) \sin(2\omega+2M_0)\\
	\sin \Omega \cos(2\omega+2M_0)+\cos~I \cdot \cos(\Omega)\sin(2\omega+2M_0)\\
	\sin~I \sin(2\omega+2M_0) 
	\end{array}
	\right]\\
	\nonumber \\
	 {\bf B_{6}}&=&- \frac{3\, d \cos~\theta_p \sin~I }{2 \, a^{3}}  \left[ 
	\begin{array}{c}
	\cos \Omega \sin(2\omega+2M_0)+\cos~I \sin(\Omega)\cos(2\omega+2M_0)\\
	\sin \Omega \sin(2\omega+2M_0)-\cos~I \cos(\Omega)\cos(2\omega+2M_0)\\
	-\sin~I \cos(2\omega+2M_0) 
	\end{array}
	\right]\\
	\nonumber \\
	 {\bf B_{7}}&=& - \frac{3\, d \sin~\theta_p }{8 \, a^{3}}   \left[ 
	\begin{array}{c}
	(1-\cos(2I)) \sin(2 \Omega-\phi_p)-\sin(\phi_p) \left(\frac{1}{3}+\cos(2I) \right)\\
	-(1-\cos(2I)) \cos(2 \Omega-\phi_p)+\cos(\phi_p) \left(\frac{1}{3}+\cos(2I) \right)\\
	2 \sin(2I) \cos(\Omega-\phi_p)
	\end{array}
	\right]\\
	\nonumber \\
	 {\bf B_{8}}&=&- \frac{3\, d \sin~\theta_p }{8 \, a^{3}}    \left[ 
	\begin{array}{c}
	(1-\cos(2I)) \cos(2 \Omega-\phi_p)+\cos(\phi_p) \left(\frac{1}{3}+\cos(2I) \right)\\
	(1-\cos(2I)) \sin(2 \Omega-\phi_p)+\sin(\phi_p) \left(\frac{1}{3}+\cos(2I) \right)\\
	-2 \sin(2I)\sin(\Omega-\phi_p)\\
	\end{array}
	\right].
	\end{eqnarray} 
}
\pagebreak
\bibliographystyle{plainnat}
\bibliography{larase,art2}

\begin{thebibliography}{60}
\providecommand{\natexlab}[1]{#1}
\providecommand{\url}[1]{\texttt{#1}}
\expandafter\ifx\csname urlstyle\endcsname\relax
  \providecommand{\doi}[1]{doi: #1}\else
  \providecommand{\doi}{doi: \begingroup \urlstyle{rm}\Url}\fi

\bibitem[{Afonso} et~al.(1989){Afonso}, {Barlier}, {Mignard}, {Carpino}, and
  {Farinella}]{1989AnGeo...7..501A}
G.~{Afonso}, F.~{Barlier}, F.~{Mignard}, M.~{Carpino}, and P.~{Farinella}.
\newblock {Orbital effects of LAGEOS seasons and eclipses}.
\newblock \emph{Ann. Geophysicae}, 7:\penalty0 501--514, October 1989.

\bibitem[{Andres} et~al.(2002){Andres}, {Noomen}, {Bianco}, {Currie}, and
  {Otsubo}]{2002EGSGA..27.2994A}
J.~I. {Andres}, R.~{Noomen}, G.~{Bianco}, D.~{Currie}, and T.~{Otsubo}.
\newblock {The Spin Axis Behavior of The Lageos Satellites}.
\newblock In A.~{Tzanis}, editor, \emph{EGS General Assembly Conference
  Abstracts}, volume~27 of \emph{EGS General Assembly Conference Abstracts},
  page 2994, 2002.

\bibitem[{Andr{\'e}s} et~al.(2003){Andr{\'e}s}, {Otsubo}, {Noomen}, {Bianco},
  and {Currie}]{2003EAEJA.....5951A}
J.~I. {Andr{\'e}s}, T.~{Otsubo}, R.~{Noomen}, G.~{Bianco}, and D.~{Currie}.
\newblock {Observing and modelling the LAGEOS spin axis behaviour}.
\newblock In \emph{EGS - AGU - EUG Joint Assembly}, page 5951, April 2003.

\bibitem[{Andr{\'e}s} et~al.(2004){Andr{\'e}s}, {Noomen}, {Bianco}, {Currie},
  and {Otsubo}]{2004JGRB..10906403A}
J.~I. {Andr{\'e}s}, R.~{Noomen}, G.~{Bianco}, D.~G. {Currie}, and T.~{Otsubo}.
\newblock {Spin axis behavior of the LAGEOS satellites}.
\newblock \emph{J. Geophys. Res.}, 109\penalty0 (B18):\penalty0 6403, June
  2004.
\newblock \doi{10.1029/2003JB002692}.

\bibitem[{Andr{\'e}s de la Fuente}(2007)]{2007Andres}
J.~I. {Andr{\'e}s de la Fuente}.
\newblock \emph{{Enhanced Modelling of LAGEOS Non-Gravitational
  Perturbations}}.
\newblock PhD thesis, Delft University Press, Sieca Repro, Turbineweg 20, 2627
  BP Delft, The Netherlands, 2007.

\bibitem[{Andr{\'e}s de la Fuente} et~al.(2006){Andr{\'e}s de la Fuente},
  {Noomen}, and {Vecellio None}]{2006Andres}
J.~I. {Andr{\'e}s de la Fuente}, R.~{Noomen}, and S.~{Vecellio None}.
\newblock {Numerical simulations of the LAGEOS thermal behavior and thermal
  accelerations}.
\newblock \emph{J. Geophys. Res.}, III:\penalty0 B09406, 2006.
\newblock \doi{10.1029/2005JB003928}.

\bibitem[{Avizonis}(1997)]{1997PhDT........14A}
Jr. P.~V. {Avizonis}.
\newblock \emph{{Remote Sensing of the Lageos-I Spin-Axis and Image Processing
  for Advanced Optical Systems}}.
\newblock PhD thesis, UNIVERSITY OF MARYLAND COLLEGE PARK, December 1997.

\bibitem[{Beletskii}(1966)]{Bele}
V.~V. {Beletskii}.
\newblock {Motion of an artificial satellite about its center of mass}.
\newblock Technical Report TT-67-51366 NASA-TT-F-429, NASA, 1966.

\bibitem[{Bertotti} and {Iess}(1991)]{1991JGR....96.2431B}
B.~{Bertotti} and L.~{Iess}.
\newblock {The rotation of Lageos}.
\newblock \emph{J. Geophys. Res.}, 96:\penalty0 2431--2440, February 1991.
\newblock \doi{10.1029/90JB01949}.

\bibitem[{Bianco} et~al.(1998){Bianco}, {Devoti}, {Fermi}, {Luceri},
  {Rutigliano}, and {Sciarretta}]{1998P&SS...46.1633B}
G.~{Bianco}, R.~{Devoti}, M.~{Fermi}, V.~{Luceri}, P.~{Rutigliano}, and
  C.~{Sciarretta}.
\newblock {Estimation of low degree geopotential coefficients using SLR data}.
\newblock \emph{Planetary and Space Science}, 46:\penalty0 1633--1638, December
  1998.
\newblock \doi{10.1016/S0032-0633(97)00215-8}.

\bibitem[Bianco et~al.(2001)Bianco, Chersich, Devoti, Luceri, and
  Selden]{GRL:GRL14288}
G.~Bianco, M.~Chersich, R.~Devoti, V.~Luceri, and M.~Selden.
\newblock Measurement of lageos-2 rotation by satellite laser ranging
  observations.
\newblock \emph{Geophysical Research Letters}, 28\penalty0 (10):\penalty0
  2113--2116, 2001.
\newblock ISSN 1944-8007.
\newblock \doi{10.1029/2000GL012435}.

\bibitem[Bidinosti et~al.(2007)Bidinosti, Chapple, and Hayden]{CMR:CMR20090}
C.P. Bidinosti, E.M. Chapple, and M.E. Hayden.
\newblock The sphere in a uniform rf field—revisited.
\newblock \emph{Concepts in Magnetic Resonance Part B: Magnetic Resonance
  Engineering}, 31B\penalty0 (3):\penalty0 191--202, 2007.
\newblock ISSN 1552-504X.
\newblock \doi{10.1002/cmr.b.20090}.
\newblock URL \url{http://dx.doi.org/10.1002/cmr.b.20090}.

\bibitem[{Chen} et~al.(2013){Chen}, {Wilson}, {Ries}, and
  {Tapley}]{2013GeoRL..40.2625C}
J.~L. {Chen}, C.~R. {Wilson}, J.~C. {Ries}, and B.~D. {Tapley}.
\newblock {Rapid ice melting drives Earth's pole to the east}.
\newblock \emph{Geophys. Res. Lett.}, 40:\penalty0 2625--2630, June 2013.
\newblock \doi{10.1002/grl.50552}.

\bibitem[{Ciufolini} and {Pavlis}(2004)]{2004Natur.431..958C}
I.~{Ciufolini} and E.~C. {Pavlis}.
\newblock {A confirmation of the general relativistic prediction of the
  Lense-Thirring effect}.
\newblock \emph{Nature}, 431:\penalty0 958--960, 2004.
\newblock \doi{10.1038/nature03007}.

\bibitem[{Ciufolini} et~al.(1996){Ciufolini}, {Lucchesi}, {Vespe}, and
  {Mandiello}]{1996NCimA.109..575C}
I.~{Ciufolini}, D.~{Lucchesi}, F.~{Vespe}, and A.~{Mandiello}.
\newblock {Measurement of dragging of inertial frames and gravitomagnetic field
  using laser-ranged satellites.}
\newblock \emph{Nuovo Cim. A}, 109:\penalty0 575--590, 1996.
\newblock \doi{10.1007/BF02731140}.

\bibitem[{Cohen} and {Smith}(1985)]{1985JGR....90.9217C}
S.~C. {Cohen} and D.~E. {Smith}.
\newblock {Lageos scientific results - Introduction}.
\newblock \emph{J. Geophys. Res.}, 90:\penalty0 9217--9220, Sep 1985.
\newblock \doi{10.1029/JB090iB11p09217}.

\bibitem[{Cox} and {Chao}(2002)]{2002Sci...297..831C}
C.~M. {Cox} and B.~F. {Chao}.
\newblock {Detection of a Large-Scale Mass Redistribution in the Terrestrial
  System Since 1998}.
\newblock \emph{Science}, 297:\penalty0 831--833, August 2002.
\newblock \doi{10.1126/science.1072188}.

\bibitem[{Farinella} and {Vokrouhlick{\'y}}(1996)]{1996P&SS...44.1551F}
P.~{Farinella} and D.~{Vokrouhlick{\'y}}.
\newblock {Thermal force effects on slowly rotating, spherical artificial
  satellites-I. Solar heating}.
\newblock \emph{Plan. Space Sci.}, 44:\penalty0 1551--1561, December 1996.
\newblock \doi{10.1016/S0032-0633(96)00073-6}.

\bibitem[{Farinella} et~al.(1990){Farinella}, {Nobili}, {Barlier}, and
  {Mignard}]{1990A&A...234..546F}
P.~{Farinella}, A.~M. {Nobili}, F.~{Barlier}, and F.~{Mignard}.
\newblock {Effects of thermal thrust on the node and inclination of LAGEOS}.
\newblock \emph{Astron. Astrophys.}, 234:\penalty0 546--554, August 1990.

\bibitem[{Farinella} et~al.(1996){Farinella}, {Vokrouhlicky}, and
  {Barlier}]{1996JGR...10117861F}
P.~{Farinella}, D.~{Vokrouhlicky}, and F.~{Barlier}.
\newblock {The rotation of LAGEOS and its long-term semimajor axis decay: A
  self-consistent solution}.
\newblock \emph{J. Geophys. Res.}, 101:\penalty0 17861--17872, August 1996.

\bibitem[{Goldstein} et~al.(2000){Goldstein}, {Poole}, and {Safko}]{Goldstein}
E.~{Goldstein}, C.~{Poole}, and J.~{Safko}.
\newblock \emph{Classical mechanics}.
\newblock Addison Wesley, Boston, 2000.

\bibitem[{Habib} et~al.(1994){Habib}, {Holz}, {Kheyfets}, {Matzner}, {Miller},
  and {Tolman}]{1994PhRvD..50.6068H}
S.~{Habib}, D.~E. {Holz}, A.~{Kheyfets}, R.~A. {Matzner}, W.~A. {Miller}, and
  B.~W. {Tolman}.
\newblock {Spin dynamics of the LAGEOS satellite in support of a measurement of
  the Earth's gravitomagnetism}.
\newblock \emph{Phys. Rev. D}, 50:\penalty0 6068--6079, November 1994.
\newblock \doi{10.1103/PhysRevD.50.6068}.

\bibitem[Hayes(1964)]{hayes_1964}
Arthur~F. Hayes.
\newblock Torque on a spinning hollow sphere in a uniform alternating magnetic
  field.
\newblock \emph{Aerospace and Navigational Electronics, IEEE Transactions on},
  ANE-11\penalty0 (2):\penalty0 122--127, June 1964.
\newblock ISSN 0096-1957.
\newblock \doi{10.1109/TANE.1964.4502174}.

\bibitem[{Kucharski} et~al.(2012){Kucharski}, {Otsubo }, {Kirchner}, and
  {Bianco }]{2012AdSpR..50.1473K}
D.~{Kucharski}, T.~{Otsubo }, G.~{Kirchner}, and G.~{Bianco }.
\newblock {Spin rate and spin axis orientation of LARES spectrally determined
  from Satellite Laser Ranging data}.
\newblock \emph{Adv. Space Res.}, 50:\penalty0 1473--1477, December 2012.
\newblock \doi{10.1016/j.asr.2012.07.018}.

\bibitem[{Kucharski} et~al.(2013){Kucharski}, {Lim}, {Kirchner}, and
  {Hwang}]{2013AdSpR..52.1332K}
D.~{Kucharski}, H.-C. {Lim}, G.~{Kirchner}, and J.-Y. {Hwang}.
\newblock {Spin parameters of LAGEOS-1 and LAGEOS-2 spectrally determined from
  Satellite Laser Ranging data}.
\newblock \emph{Adv.Space Res.}, 52:\penalty0 1332--1338, October 2013.
\newblock \doi{10.1016/j.asr.2013.07.007}.

\bibitem[Kucharski et~al.(2014)Kucharski, Lim, Kirchner, Otsubo, Bianco, and
  Hwang]{6575147}
D.~Kucharski, Hyung-Chul Lim, G.~Kirchner, T.~Otsubo, G.~Bianco, and Joo-Yeon
  Hwang.
\newblock Spin axis precession of lares measured by satellite laser ranging.
\newblock \emph{Geoscience and Remote Sensing Letters, IEEE}, 11\penalty0
  (3):\penalty0 646--650, March 2014.
\newblock ISSN 1545-598X.
\newblock \doi{10.1109/LGRS.2013.2273561}.

\bibitem[Landau and Lifshitz(1960)]{landau_1960}
L~Landau and E~Lifshitz.
\newblock \emph{Electrodynamics of continuous media}.
\newblock Pergamon Press, Oxford-New York, 1960.

\bibitem[{Lemoine} et~al.(1998){Lemoine}, {Kenyon}, {Factor}, {Trimmer},
  {Pavlis}, {Chinn}, {Cox}, {Klosko}, {Luthcke}, {Torrence}, {Wang},
  {Williamson}, {Pavlis}, {Rapp}, and {Olson}]{1998Lemoine}
F.~G. {Lemoine}, S.C. {Kenyon}, J.~K. {Factor}, R.~G. {Trimmer}, N.~K.
  {Pavlis}, D.~S. {Chinn}, C.~M. {Cox}, S.~M. {Klosko}, S.~B. {Luthcke}, M.~H.
  {Torrence}, Y.~M. {Wang}, R.~G. {Williamson}, E.~C. {Pavlis}, R.~H. {Rapp},
  and T.~R. {Olson}.
\newblock {The Development of the Joint NASA GSFC and the National Imagery and
  Mapping Agency (NIMA) Geopotential Model EGM96}.
\newblock Technical Paper 206861, NASA, 1998.

\bibitem[{Lucchesi}(2002)]{2002P&SS...50.1067L}
D.~M. {Lucchesi}.
\newblock {Reassessment of the error modelling of non-gravitational
  perturbations on LAGEOS II and their impact in the Lense-Thirring
  derivation-Part II}.
\newblock \emph{Plan. Space Sci.}, 50:\penalty0 1067--1100, August 2002.
\newblock \doi{10.1016/S0032-0633(02)00052-1}.

\bibitem[{Lucchesi}(2003)]{2003GeoRL..30.1957L}
D.~M. {Lucchesi}.
\newblock {The asymmetric reflectivity effect on the LAGEOS satellites and the
  germanium retroreflectors}.
\newblock \emph{Geophys. Res. Lett.}, 30:\penalty0 1957, September 2003.
\newblock \doi{10.1029/2003GL017741}.

\bibitem[{Lucchesi}(2004)]{2004CeMDA..88..269L}
D.~M. {Lucchesi}.
\newblock {LAGEOS Satellites Germanium Cube-Corner-Retroreflectors and the
  Asymmetric Reflectivity Effect}.
\newblock \emph{Celest. Mech. Dyn. Astron.}, 88:\penalty0 269--291, March 2004.
\newblock \doi{10.1023/B:CELE.0000017171.78328.f1}.

\bibitem[{Lucchesi}(2007)]{2007AdSpR..39..324L}
D.~M. {Lucchesi}.
\newblock {The Lense Thirring effect measurement and LAGEOS satellites orbit
  analysis with the new gravity field model from the CHAMP mission}.
\newblock \emph{Adv. Space Res.}, 39:\penalty0 324--332, 2007.
\newblock \doi{10.1016/j.asr.2006.10.012}.

\bibitem[{Lucchesi} and {Peron }(2014)]{2014PhRvD..89h2002L}
D.~M. {Lucchesi} and R.~{Peron }.
\newblock {LAGEOS II pericenter general relativistic precession (1993-2005):
  Error budget and constraints in gravitational physics}.
\newblock \emph{Phys. Rev. D}, 89\penalty0 (8):\penalty0 082002, April 2014.
\newblock \doi{10.1103/PhysRevD.89.082002}.

\bibitem[{Lucchesi} and {Peron}(2010)]{2010PhRvL.105w1103L}
D.~M. {Lucchesi} and R.~{Peron}.
\newblock {Accurate Measurement in the Field of the Earth of the
  General-Relativistic Precession of the LAGEOS II Pericenter and New
  Constraints on Non-Newtonian Gravity}.
\newblock \emph{Phys. Rev. Lett.}, 105\penalty0 (23):\penalty0 231103, December
  2010.
\newblock \doi{10.1103/PhysRevLett.105.231103}.

\bibitem[{Lucchesi} et~al.(2004){Lucchesi}, {Ciufolini}, {Andr{\'e}s},
  {Pavlis}, {Peron}, {Noomen}, and {Currie}]{2004P&SS...52..699L}
D.~M. {Lucchesi}, I.~{Ciufolini}, J.~I. {Andr{\'e}s}, E.~C. {Pavlis},
  R.~{Peron}, R.~{Noomen}, and D.~G. {Currie}.
\newblock {LAGEOS II perigee rate and eccentricity vector excitations residuals
  and the Yarkovsky-Schach effect}.
\newblock \emph{Plan. Space Sci.}, 52:\penalty0 699--710, July 2004.
\newblock \doi{10.1016/j.pss.2004.01.007}.

\bibitem[{Lucchesi} et~al.(2015){Lucchesi}, {Anselmo}, {Bassan}, {Pardini},
  {Peron}, {Pucacco}, and {Visco}]{2015CQGra..32o5012L}
D.~M. {Lucchesi}, L.~{Anselmo}, M.~{Bassan}, C.~{Pardini}, R.~{Peron},
  G.~{Pucacco}, and M.~{Visco}.
\newblock {Testing the gravitational interaction in the field of the Earth via
  satellite laser ranging and the Laser Ranged Satellites Experiment (LARASE)}.
\newblock \emph{Classical and Quantum Gravity}, 32\penalty0 (15):\penalty0
  155012, August 2015.
\newblock \doi{10.1088/0264-9381/32/15/155012}.

\bibitem[{Lucchesi} et~al.(2016){Lucchesi}, {Magnafico}, {Peron}, {Visco},
  {Anselmo}, {Pardini}, {Bassan}, {Pucacco}, and {Stanga}]{7573270}
D.M. {Lucchesi}, C.~{Magnafico}, R.~{Peron}, M.~{Visco}, L.~{Anselmo},
  C.~{Pardini}, M.~{Bassan}, G.~{Pucacco}, and R.~{Stanga}.
\newblock {Measurements of general relativity precessions in the field of the
  Earth with laser-ranged satellites and the LARASE program}.
\newblock In \emph{{2016 IEEE Metrology for Aerospace (MetroAeroSpace)}}, pages
  522--529, 2016.
\newblock \doi{10.1109/MetroAeroSpace.2016.7573270}.

\bibitem[{M{\'e}tris} et~al.(1997){M{\'e}tris}, {Vokrouhlick{\'y}}, {Ries}, and
  {Eanes}]{1997JGR...102.2711M}
G.~{M{\'e}tris}, D.~{Vokrouhlick{\'y}}, J.~C. {Ries}, and R.~J. {Eanes}.
\newblock {Nongravitational effects and the LAGEOS eccentricity excitations}.
\newblock \emph{J. Geophys. Res.}, 102:\penalty0 2711--2729, February 1997.
\newblock \doi{10.1029/96JB03186}.

\bibitem[{M{\'e}tris} et~al.(1999){M{\'e}tris}, {Vokrouhlick{\'y}}, {Ries },
  and {Eanes }]{1999AdSpR..23..721M}
G.~{M{\'e}tris}, D.~{Vokrouhlick{\'y}}, J.~C. {Ries }, and R.~J. {Eanes }.
\newblock {LAGEOS Spin Axis and Non-gravitational Excitations of its Orbit}.
\newblock \emph{Adv. Space Res.}, 23:\penalty0 721--725, 1999.
\newblock \doi{10.1016/S0273-1177(99)00142-8}.

\bibitem[{{Montenbruck}, O. and {Gill}, E.}(2005)]{montenbruk}
{{Montenbruck}, O. and {Gill}, E.}
\newblock \emph{SatelliteOrbits- models, methods and Application}.
\newblock Springer, Berlin, 2005.

\bibitem[{Nakagawa} et~al.(2000){Nakagawa}, {Ishii}, {Tsuruda}, {Hayakawa}, and
  {Mukai}]{2000Nakagawa}
T.~{Nakagawa}, T.~{Ishii}, K.~{Tsuruda}, H.~{Hayakawa}, and T.~{Mukai}.
\newblock Net current density of photoelectrons emitted from the surface of the
  geotail spacecraft.
\newblock \emph{Earth Planets Space}, 52:\penalty0 283--292, 2000.
\newblock \doi{10.1186/BF03351637}.

\bibitem[{Otsubo} et~al.(2004){Otsubo}, {Sherwood}, {Gibbs}, and
  {Wood}]{2004ITGRS..42..202O}
T.~{Otsubo}, R.~A. {Sherwood}, P.~{Gibbs}, and R.~{Wood}.
\newblock {Spin Motion and Orientation of LAGEOS-2 From Photometric
  Observation}.
\newblock \emph{IEEE Transactions on Geoscience and Remote Sensing},
  42:\penalty0 202--208, January 2004.
\newblock \doi{10.1109/TGRS.2003.817191}.

\bibitem[{Paolozzi} and {Ciufolini}(2013)]{2013AcAau..91..313P}
A.~{Paolozzi} and I.~{Ciufolini}.
\newblock {LARES successfully launched in orbit: Satellite and mission
  description}.
\newblock \emph{Acta Astronautica}, 91:\penalty0 313--321, October 2013.
\newblock \doi{10.1016/j.actaastro.2013.05.011}.

\bibitem[{Pardini} et~al.(2017){Pardini}, {Anselmo}, {Lucchesi}, and
  {Peron}]{2017Pardini}
C.~{Pardini}, L.~{Anselmo}, D.M. {Lucchesi}, and R.~{Peron}.
\newblock On the secular decay of the lares semi-major axis.
\newblock \emph{Acta Astronautica}, 140:\penalty0 469--477, 2017.
\newblock \doi{10.1016/j.actaastro.2017.09.012}.

\bibitem[{Pavlis} and {al.}(1998)]{1998pavlis}
D.~E. {Pavlis} and {al.}
\newblock \emph{{GEODYN II Operations Manual}}.
\newblock NASA GSFC, 1998.

\bibitem[{Pearlman} et~al.(2002){Pearlman}, {Degnan}, and
  {Bosworth}]{2002AdSpR..30..135P}
M.~R. {Pearlman}, J.~J. {Degnan}, and J.~M. {Bosworth}.
\newblock {The International Laser Ranging Service}.
\newblock \emph{Adv. Space Res.}, 30:\penalty0 135--143, 2002.

\bibitem[{Putney} et~al.(1990){Putney}, {Kolenkiewcz}, {Smith}, {Dunn}, and
  {Torrence}]{1990Putney}
B.~{Putney}, K.~{Kolenkiewcz}, D.~{Smith}, P.~{Dunn}, and M.H. {Torrence}.
\newblock Precision orbit determination at the nasa goddard space flight
  center.
\newblock \emph{Adv. Space Res.}, 10:\penalty0 197--203, 1990.
\newblock \doi{10.1016/0273-1177(90)90350-9}.

\bibitem[{Rubincam}(1984)]{1984JGR....89.1077R}
D.~P. {Rubincam}.
\newblock {Postglacial rebound observed by Lageos and the effective viscosity
  of the lower mantle}.
\newblock \emph{J. Geophys. Res.}, 89:\penalty0 1077--1087, February 1984.
\newblock \doi{10.1029/JB089iB02p01077}.

\bibitem[{Rubincam}(1988)]{1988JGR....9313805R}
D.~P. {Rubincam}.
\newblock {Yarkovsky thermal drag on LAGEOS}.
\newblock \emph{J. Geophys. Res.}, 93:\penalty0 13805--13810, November 1988.
\newblock \doi{10.1029/JB093iB11p13805}.

\bibitem[{Rubincam} et~al.(1987){Rubincam}, {Knocke}, {Taylor}, and
  {Blackwell}]{1987JGR....9211662R}
D.~P. {Rubincam}, P.~{Knocke}, V.~R. {Taylor}, and S.~{Blackwell}.
\newblock {Earth anisotropic reflection and the orbit of LAGEOS}.
\newblock \emph{J. Geophys. Res.}, 92:\penalty0 11662--11668, October 1987.
\newblock \doi{10.1029/JB092iB11p11662}.

\bibitem[{Rubincam} et~al.(1997){Rubincam}, {Currie}, and
  {Robbins}]{1997JGR...102..585R}
D.~P. {Rubincam}, D.~G. {Currie}, and J.~W. {Robbins}.
\newblock {LAGEOS I once-per-revolution force due to solar heating}.
\newblock \emph{J. Geophys. Res.}, 102:\penalty0 585--590, January 1997.
\newblock \doi{10.1029/96JB02851}.

\bibitem[{Scharroo} et~al.(1991){Scharroo}, {Wakker}, {Ambrosius}, and
  {Noomen}]{1991JGR....96..729S}
R.~{Scharroo}, K.~F. {Wakker}, B.~A.~C. {Ambrosius}, and R.~{Noomen}.
\newblock {On the along-track acceleration of the Lageos satellite}.
\newblock \emph{J. Geophys. Res.}, 96:\penalty0 729--740, January 1991.
\newblock \doi{10.1029/90JB02080}.

\bibitem[{Sinclair}(2012)]{Sinclair2012}
A.~T. {Sinclair}.
\newblock {Data Screening and Normal Point Formation --- Re--Statement of
  Herstmonceux Normal Point Recommendation}, 2012.
\newblock URL
  \url{http://ilrs.gsfc.nasa.gov/data\_and\_products/data/npt/npt\_algorithm.html}.

\bibitem[{Slabinski}(1996)]{1996CeMDA..66..131S}
V.~J. {Slabinski}.
\newblock {A Numerical Solution for Lageos Thermal Thrust: The Rapid-Spin
  Case}.
\newblock \emph{Celest. Mech. Dyn. Astron.}, 66:\penalty0 131--179, June 1996.
\newblock \doi{10.1007/BF00054962}.

\bibitem[{Smith} et~al.(1990){Smith}, {Kolenkiewicz}, {Dunn}, {Robbins},
  {Torrence}, {Klosko}, {Williamson}, {Pavlis}, and
  {Douglas}]{1990JGR....9522013S}
D.~E. {Smith}, R.~{Kolenkiewicz}, P.~J. {Dunn}, J.~W. {Robbins}, M.~H.
  {Torrence}, S.~M. {Klosko}, R.~G. {Williamson}, E.~C. {Pavlis}, and N.~B.
  {Douglas}.
\newblock {Tectonic motion and deformation from satellite laser ranging to
  Lageos}.
\newblock \emph{J. Geophys. Res.}, 95:\penalty0 22013--22041, December 1990.
\newblock \doi{10.1029/JB095iB13p22013}.

\bibitem[{Sullivan}(1980)]{1980SPIE..227..148S}
L.~J. {Sullivan}.
\newblock {Infrared coherent radar}.
\newblock In T.~S. {Hartwick}, editor, \emph{{CO2 laser devices and
  applications}}, volume 227 of \emph{{Society of Photo-Optical Instrumentation
  Engineers (SPIE) Conference Series}}, pages 148--161, January 1980.

\bibitem[{Visco} and {Lucchesi}(2016)]{2016AdSpR..57.1928V}
M.~{Visco} and D.~M. {Lucchesi}.
\newblock {Review and critical analysis of mass and moments of inertia of the
  LAGEOS and LAGEOS II satellites for the LARASE program}.
\newblock \emph{Advances in Space Research}, 57:\penalty0 1928--1938, May 2016.
\newblock \doi{10.1016/j.asr.2016.02.006}.

\bibitem[{Vokrouhlick{\'y}}(1996)]{1996GeoRL..23.3079V}
D.~{Vokrouhlick{\'y}}.
\newblock {Non-gravitational effects and LAGEOS' rotation}.
\newblock \emph{Geophys. Res. Lett.}, 23:\penalty0 3079--3082, 1996.
\newblock \doi{10.1029/96GL03025}.

\bibitem[{Williams}(2002)]{2002PhDT.......176W}
S.~E. {Williams}.
\newblock \emph{{The Lageos Satellite: A Comprehensive Spin Model and
  Analysis}}.
\newblock PhD thesis, NCSU PhD Dissertation, pp.~i-xii, 1-252, 2002, December
  2002.

\bibitem[{Yoder} et~al.(1983){Yoder}, {Williams}, {Dickey}, {Schutz}, {Eanes},
  and {Tapley}]{1983Natur.303..757Y}
C.~F. {Yoder}, J.~G. {Williams}, J.~O. {Dickey}, B.~E. {Schutz}, R.~J. {Eanes},
  and B.~D. {Tapley}.
\newblock {Secular variation of earth's gravitational harmonic J2 coefficient
  from Lageos and nontidal acceleration of earth rotation}.
\newblock \emph{Nature}, 303:\penalty0 757--762, June 1983.
\newblock \doi{10.1038/303757a0}.

\end{thebibliography}

\end{document}